\documentclass[10pt]{iopart}

\usepackage{cite}
\usepackage{xcolor}
\usepackage{hyperref}
\hypersetup{
    colorlinks=true,
    linkcolor=blue,
    citecolor=blue,
    urlcolor=blue,
    pdfborder={0 0 0},
    breaklinks=true, 
}

\expandafter\let\csname equation*\endcsname=\relax 
\expandafter\let\csname endequation*\endcsname=\relax

\usepackage{graphicx,subcaption,float,mathrsfs,amsmath,amssymb,amsthm,physics}

\begin{document}

\title{Path integral derivation of the thermofield double state in causal diamonds}

\author{Abhijit Chakraborty$^1$, Carlos R. Ordóñez$^2$\\and Gustavo Valdivia-Mera$^2$\footnote[1]{Author to whom any correspondence should be addressed.}}
\address{$^1$ Institute for Quantum Computing, University of Waterloo, Waterloo,\\ON, N2L 3G1, Canada}
\address{$^2$ Department of Physics, University of Houston, Houston,\\Texas 77204-5005, USA}
\eads{\mailto{abhijit.chakraborty@uwaterloo.ca}, \mailto{cordonez@central.uh.edu}, \mailto{gvaldiviamera@uh.edu}}
\vspace{10pt}
\begin{indented}
\item[]May 2024
\end{indented}

\begin{abstract}
In this article, we adopt the framework developed by R. Laflamme in \textit{Physica A}, \textbf{158}, pp. 58-63 (1989) to analyze the path integral of a massless -- conformally invariant -- scalar field defined on a causal diamond of size $2\alpha$ in 1+1 dimensions. By examining the Euclidean geometry of the causal diamond, we establish that its structure is conformally related to the cylinder $S^{1}_{\beta} \otimes \mathbb{R}$, where the Euclidean time coordinate $\tau$ has a periodicity of $\beta$. This property, along with the conformal symmetry of the fields, allows us to identify the connection between the thermofield double (TFD) state of causal diamonds and the Euclidean path integral defined on the two disconnected manifolds of the cylinder. Furthermore, we demonstrate that the temperature of the TFD state, derived from the conditions in the Euclidean geometry and analytically calculated, coincides with the temperature of the causal diamond known in the literature. This derivation highlights the universality of the connection between the Euclidean path integral formalism and the TFD state of the causal diamond, as well as it further establishes causal diamonds as a model that exhibits all desired properties of a system exhibiting the Unruh effect.

\end{abstract}

\section{Introduction}

Stephen Hawking, in his seminal work in 1975, showed that black holes can emit thermal radiation if one considers quantum effects \cite{Hawking:1974rv}. Hawking's results \cite{Hawking:1974rv, Hawking:1975vcx,Hawking:1976de} were subsequently corroborated by a series of papers by Unruh, Fulling, Davies, Parker, and Wald \cite{Unruh:1976db,Davies:1976ei,Davies:1976hi,Davies:1977bgr, Davies:1974th,Wald:1975kc,Parker:1975jm}, which flourished into a new field of research on quantum effects in curved spacetime.\\

Around the same time, Umezawa and Takahashi, motivated by many-body physics, developed the thermofield dynamics formalism, constructing a temperature-dependent vacuum state $\ket{0(\beta)}$, known as the thermofield double state (TFD) \cite{Takahasi:1974zn,Takahashi:1996zn,Umezawa:1982nv,Matsumoto:1982ry}. This state was designed in a way such that the thermal ensemble average of any observable can be written as the expectation value of the observable with respect to this temperature-dependent vacuum: \(\bra{0(\beta)}F\ket{0(\beta)}=Z^{-1}(\beta)\sum_n e^{-\beta E_n}\bra{n}F\ket{n}\), where $F$ represents an observable, $E_n$ are energy eigenvalues, and $Z\qty(\beta)$ is the partition function. To achieve this, they introduced a fictitious system mirroring the original one, showing that the `thermal' vacuum state can be expressed as \(\ket{0(\beta)}=Z^{-1/2}(\beta)\sum_n e^{-\beta E_n/2}\ket{n,\widetilde{n}}\), where \(\ket{\widetilde{n}}\) is the Hamiltonian eigenbasis of the fictitious system.\\

Shortly after Umezawa and Takahashi's introduction of thermofield dynamics, Israel showed that one can interpret the vacuum state for a scalar field theory defined on extensions of Schwarschild and Rindler spacetimes -- Kruskal and Minkowski, respectively -- as the Umezawa-Takahashi's thermofield double (TFD) state \cite{Israel:1976ur}. Time-reversed copies of Rindler and Schwarzschild geometries can be identified within these extended spacetimes. It is precisely on these copies that the fictitious fields are defined. Since an observer restricted to the Rindler and Schwarzschild spacetime will only have access to the original system, the degrees of freedom of the fictitious system need to be averaged over, which gives rise to the thermal behavior of the Mikowski or Kruskal vacuum. In other words, the emergence of the thermal behavior in the TFD state takes place when the observation of particle modes is restricted due to the presence of horizons. The temperature of the TFD state is usually determined by the intrinsic parameters of the system, such as surface gravity or acceleration in the Schwarzschild and Rindler case, respectively.\\

In a series of beautiful papers, Laflamme later showed that the TFD state can be obtained through the path integral approach \cite{Laflamme:1989cm,Laflamme:1988wg}. In this context, the field's boundary conditions are set on Euclidean sections obtained by unwrapping the original manifold, where the identification of the Euclidean time $\tau\sim\tau+\beta$ takes place. Additionally, the periodicity $\beta$ holds a deeper significance as it bridges the gap between geometry and thermodynamics. When working within Euclidean sections and developing the field theory, 
this value precisely corresponds to the inverse of the temperature of the ensemble described by the TFD state.\\

In this article, we will apply Laflamme's elegant methodology to the causal diamond (CD) \cite{Martinetti:2002sz,Martinetti:2008ja}. Unlike the Unruh-Davis effect, where observers experience acceleration, here the observers are stationary but have a finite lifetime, leading to past and future horizons that define the CD. These horizons generate a thermal spectrum, resulting in the physical system manifesting as a mixed quantum state. Causal diamonds offer unique insights into Unruh-like radiation, emphasizing the role of the horizon in thermality without requiring acceleration. This effect is more experimentally accessible and has garnered recent interest due to its relevance to black hole physics and quantum many-body theory \cite{Jacobson:2018ahi,Jacobson:2022gmo,Wang:2019zhr,Arzano:2020thh,2661076,Chandrasekaran:2019ewn,deBoer:2016pqk,Gibbons:2007nm,Foo:2020rsy,AndradeeSilva:2022iic,Su:2015oys}. Our goals in applying the method of reference \cite{Laflamme:1988wg} to the causal diamond are: (a) we want to attract the attention of the community to this less-utilized Euclidean path-integral formalism developed by Laflamme and show its universality; (b) by using Laflamme's method, we move one step closer to showing that all the different methods used for explaining the thermality in Rindler geometry apply to causal diamonds as well, without the need of an accelerated observer. This will help establish causal diamonds as a more fundamental setup to understand the role of horizon in producing thermal effects in a vacuum, as compared to the Rindler geometry.\\

Our article is organized as follows. In section \ref{sec2}, we provide a brief summary of the framework used by Laflamme to set up our calculations. In section \ref{sec3}, we analyze the Euclidean action of a conformally invariant quantum field theory in \(1+1\) dimensions, where a massless scalar field is defined on a causal diamond background. Through suitable transformations, we adapt this action into the form of a harmonic oscillator, which has an exact solution in the Euclidean path integral, leading us to the TFD formalism. In section \ref{tfdss}, we compute the density matrix of the original physical system and demonstrate that the initially imposed geometric conditions enable us to determine the equilibrium temperature of the thermal bath. Finally, we conclude with a discussion of this framework in section \ref{concl}.

\begin{figure}[t]
    \centering
    \begin{subfigure}{0.4\textwidth}
        \centering
        \includegraphics[width=\textwidth]{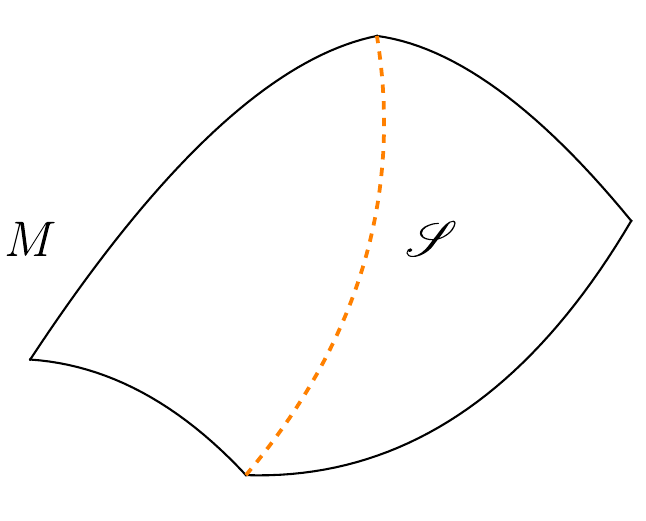}
        \caption{}
        \label{fig:subfig1}
    \end{subfigure}
    \hspace{0.1cm}
    \begin{subfigure}{0.5\textwidth}
        \centering
        \begin{subfigure}{0.49\textwidth}
            \centering
            \includegraphics[width=\textwidth]{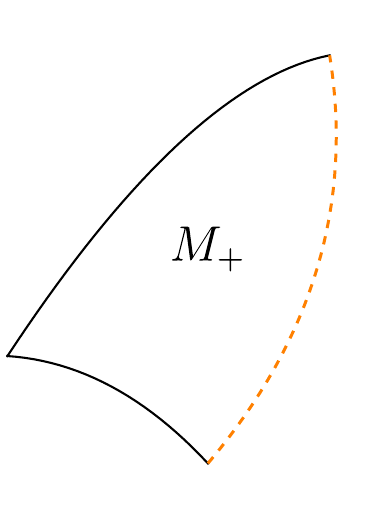}
        \end{subfigure}
        \begin{subfigure}{0.49\textwidth}
            \centering
            \includegraphics[width=\textwidth]{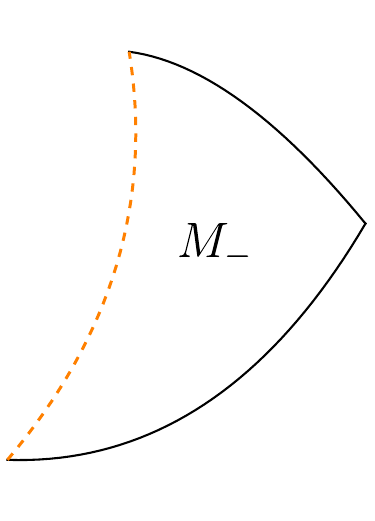}
        \end{subfigure}
        \caption{}
        \label{fig:subfig2y3}
    \end{subfigure}
    \caption{(a) Manifold \(M\) and surface \(\mathscr{S}\). (b) Submanifolds \(M_+\) and \(M_-\) obtained through the cut on the surface \(\mathscr{S}\).}
    \label{fig1}
\end{figure}

\section{Brief review of the geometric interpretation of a TFD state}\label{sec2}
In this section, we provide a brief overview of the method used by Laflamme in Refs.~\cite{Laflamme:1989cm,Laflamme:1988wg} to elucidate the connection between geometry and the thermofield double structure of the vacuum state.

\subsection{The Euclidean formalism}\label{appa}

In quantum field theory on Euclidean manifolds, the basic assumption is that the probability density for a certain configuration of the fields is proportional to $e^{-S_E[\Phi]}$, where $S_E[\Phi]$ is the Euclidean action of the field $\Phi$. Thus, the probability that the field possesses a certain property $A$, modulo gauge subtleties, is given by
\begin{equation}
    P(A) = \int_\mathscr{C} D\Phi \;\Pi(A) \,e^{-S_E\qty[\Phi]}.
\end{equation}
Here, \(\Pi(A)\) is equal to 1 if the field possesses the mentioned property, or 0 otherwise, and \(\mathscr{C}\) represents the class of regular fields defined on the Euclidean manifold \(M\). By regular fields, we mean those that do not contain singularities at any point in spacetime and whose derivatives are smooth everywhere. In the method developed by Laflamme, the property $A$ is defined as the existence of a surface $\mathscr{S}$ that divides $M$ into two parts, $M_\pm$ (see figure \ref{fig1}) connected only at $\mathscr{S}$, such that the field configuration on that surface is given by $\Tilde{\Phi}$. Moreover, given that both submanifolds \(M_\pm\) share the property $A$, \(P(A)\) will receive contributions from both of them via the action $S_E\qty[\Phi]$. In other words, the probability can be factorized into the product
\begin{equation}
    P(A)=\Psi_+(A) \Psi_-(A),\label{probiniti}
\end{equation}
where \(\Psi_\pm(A)\) is the Euclidean path integral of the field \(\Phi\) over the manifold \(M_\pm\). Similarly, we know that this path integral can be expressed as the transition amplitude between states defined on the boundaries of \(M_\pm\) (see \cite{feynman2010quantum}), where the propagation between these states is driven by an evolution operator along a specific axis of the Euclidean plane.

\begin{figure}[t]
    \centering
    \begin{subfigure}{0.25\textwidth}
        \includegraphics[width=\linewidth]{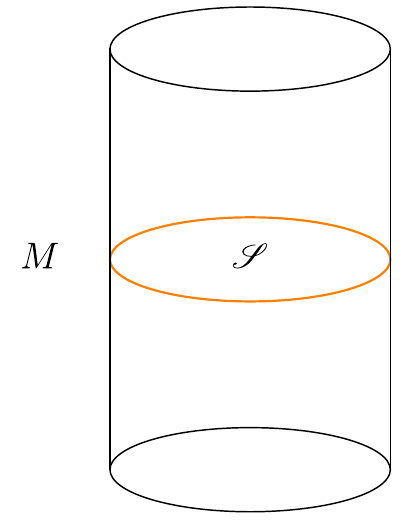}
\captionsetup{justification=centering}
        \caption{}
        \label{fig20}
    \end{subfigure}
\hspace{1cm}
    \begin{subfigure}{0.55\textwidth}
        \includegraphics[width=\linewidth]{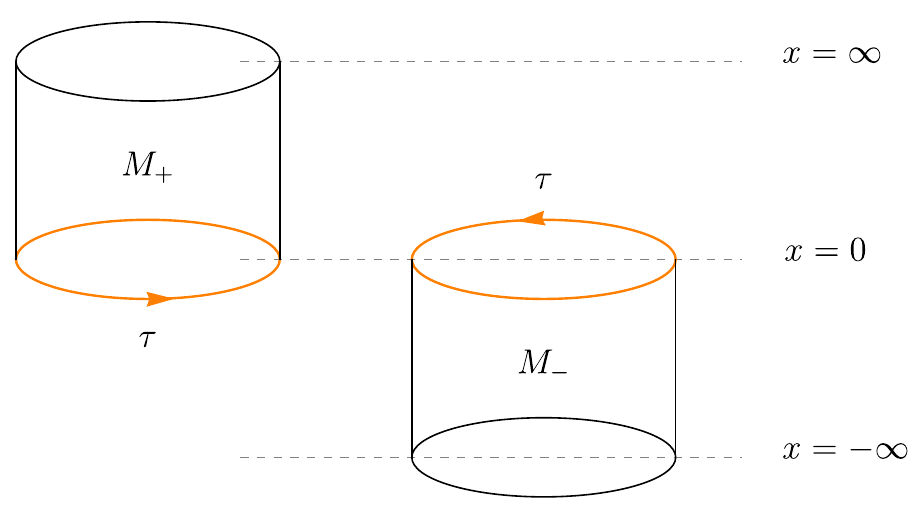}
\captionsetup{justification=centering}
        \caption{}
        \label{fig21}
    \end{subfigure}
\hspace{1cm}
    \begin{subfigure}{0.25\textwidth}
        \includegraphics[width=\linewidth]{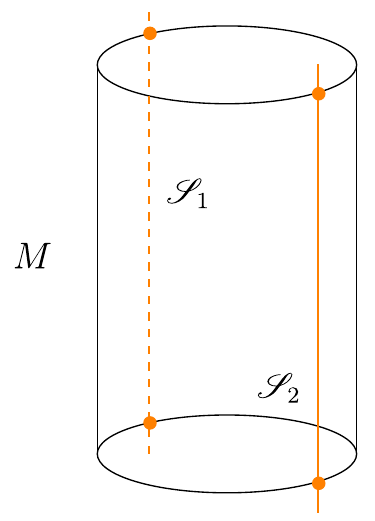}
\captionsetup{justification=centering} 
        \caption{}
        \label{fig31}
    \end{subfigure}
    \hspace{.5cm}
    \begin{subfigure}{0.25\textwidth}
        \includegraphics[width=\linewidth]{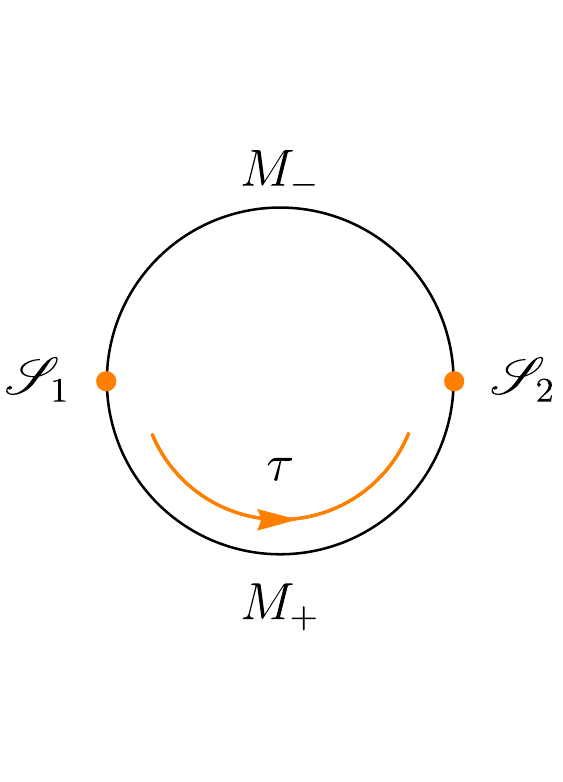}
\captionsetup{justification=centering}
        \caption{}
        \label{fig32}
    \end{subfigure}
    \hspace{.5cm}
    \begin{subfigure}{0.25\textwidth}
        \includegraphics[width=\linewidth]{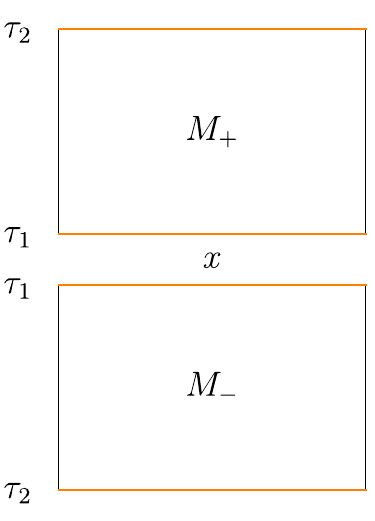}
\captionsetup{justification=centering}
        \caption{}
        \label{fig33}
    \end{subfigure}    
    \caption{(a) Manifold $M$ given by the cylinder $S^1_\beta\otimes \mathbb{R}$ and the surface $\mathscr{S}= x_0 = 0$. (b) Submanifolds \(M_+\) and \(M_-\) obtained through the cut on the surface \(\mathscr{S} = x_0 = 0\). (c) Manifold $M$ given by the cylinder $S^1\otimes \mathbb{R}$ and the surfaces $\mathscr{S}_1=\tau_1$ and $\mathscr{S}_2=\tau_2$. (d) Transversal view of the cylinder. (e) Submanifolds \(M_+\) and \(M_-\) obtained through the cuts on the surfaces \(\qty{\mathscr{S}_1 = \tau_1, \mathscr{S}_2 = \tau_2}\).}
    \label{fig:submanifolds}
\end{figure}

\subsection{The cylinder}\label{secyl}
In Laflamme's method, the manifold \( M \) is given by the cylinder \( S^{1}_{\beta} \otimes \mathbb{R} \). The Euclidean line element of this cylinder is \( ds_E^2 = d\tau^2 + dx^2 \), where \(\tau\) has a periodicity of \(\beta\). According to the method described in \cite{Laflamme:1988wg}, this manifold now needs to be divided into two submanifolds. As shown in figure \ref{fig:submanifolds}, the surface $\mathscr{S}$ dividing the manifold $M$ can be defined in two ways: (a) as a single surface cutting the cylinder into a top and bottom half, and (b) as the union of two disjointed surface $\mathscr{S} := \mathscr{S}_1 \cup \mathscr{S}_2$ that allows the cylinder to unroll into a plane. In this article, we are interested in the second construction as it allows us to elucidate the connection between the fictitious fields in the TFD states with the path integral formalism. Moreover, identifying \(x\) and \(\tau\) with the Lorentzian metric \(ds^2 = -d\sigma^2 + dx^2\) through the Wick rotation \(\tau = i\sigma\) allows us to interpret \(\tau\) as the Euclidean time coordinate. Thus, because the time interval in both submanifolds is \(\Delta \tau = \beta/2\), we obtain that the vacuum state will be described by a thermofield double structure with a finite non-zero temperature determined by \(T = \beta^{-1}\). In \ref{app:horizontal_cut_temp}, we discuss briefly why the first construction is not associated with a non-zero temperature.\\

With the choice of the second construction, the boundary conditions of the fields can be defined as \(\Phi_1\) on \(\mathscr{S}_1\) and \(\Phi_2\) on \(\mathscr{S}_2\). In this way, the Euclidean path integrals on each submanifold will be given by the following transition amplitudes (see \cite{feynman2010quantum}):
\begin{align}
    \Psi_+[\Phi_1,\Phi_2] &= \int_{\substack{\Phi_1 \\ \mathstrut \hspace{-2.5em} \mathscr{C}_+}}^{\Phi_2}
 D\Phi \, e^{-S_E[\Phi]}=\bra{\Phi_2}e^{-\frac{\beta}{2}H}\ket{\Phi_1},\label{psi+++}\\
\Psi_-[\Phi_1,\Phi_2] &= \int_{\substack{\Phi_2 \\ \mathstrut \hspace{-2.5em} \mathscr{C}_-}}^{\Phi_1} D\Phi \, e^{-S_E[\Phi]}=\bra{\Phi_1}e^{-\frac{\beta}{2}H}\ket{\Phi_2},\label{psi---}
\end{align}
where \(\mathscr{C}_\pm\) denotes the class of regular fields defined on the background geometry \(M_\pm\), and $H$ is the Hamiltonian. Additionally, as we can observe from \eqref{psi+++} and \eqref{psi---}, the transition amplitudes \(\Psi_+\) and \(\Psi_-\) are complex conjugates, that is, \(\Psi_+ = \Psi^*_-\), which makes the probabilistic interpretation of the given field configurations \(\Tilde{\Phi} = \{\Phi_1, \Phi_2\}\) on the surfaces \(\mathscr{S} = \{\mathscr{S}_1, \mathscr{S}_2\}\) mentioned in \eqref{probiniti} evident:
\begin{equation}
    P_{\mathscr{S}}[\Tilde{\Phi}] = \Psi_+[\Phi_1, \Phi_2] \Psi_-[\Phi_1, \Phi_2].
\end{equation}

\subsection{The Euclidean path integral and density matrix}\label{phinttrsam}
As we can observe from equations \eqref{psi+++} and \eqref{psi---}, there are two key aspects to address:\\

(1) The Euclidean path integral: In Laflamme's method, the Euclidean action is analyzed to identify the periodic coordinate \(\tau\), which will be associated with the hypercylinder. This identification allows us to understand how the geometric flow generated by \(\partial_\tau\) acts in the extended geometry. From this, we can define the boundary conditions in the path integral, i.e., the field configuration at the boundaries of the submanifolds \(M_\pm\). Furthermore, under certain transformations, it is possible to bring the field action into the form of a harmonic oscillator. Therefore, the path integral can be computed exactly.\\

(2) The transition amplitude: Here, the Hamiltonian operator \(H\) fulfills two fundamental roles. First, it implements the geometric flow of the vector field \(\partial_\tau\) at the level of quantum states, so that \(\Psi_+\) is determined by the propagation of the state \(\ket{\Phi_1}\) to \(\ket{\Phi_2}\) through an evolution in \(\tau\) of \(\beta/2\), and vice versa for \(\Psi_-\). Second, we can express the field configurations \(\ket{\Phi_1}\) and \(\ket{\Phi_2}\) in the eigenbasis of the Hamiltonian, \(\ket{\Phi_{1,2}} = \sum_n \bra{n}\ket{\Phi_{1,2}} \ket{n}\), since \(H\) is defined over the manifold \(M\), given that \(\tau\) is defined over the entire cylinder.\\

By evaluating the Euclidean path integral exactly for the hypercylinder $S^1_\beta \otimes \mathbb{R}$ and using the energy eigenbasis, Laflamme showed that the transition amplitude can be written as a TFD state:
\begin{equation}
    \Psi_\pm\qty[\Phi_1,\Phi_2]=\frac{1}{\sqrt{Z\qty(\beta)}}\sum_{n=0}^{\infty}e^{-\frac{\beta}{2}E_n}\varphi_n\qty[\Phi_1]\varphi_n\qty[\Phi_2],
\end{equation}
where $\varphi_n[\cdot]$ denotes the energy eigenbasis corresponding to the Hamiltonian, and $Z(\beta)$ is the partition function. For an observer on one of the surfaces (say, $\mathscr{S}_1$), the field configuration on the disconnected surface ($\mathscr{S}_2$) is `fictitious', and hence we must integrate over this fictitious field to obtain the physical density matrix from the transition amplitudes:
\begin{equation}
    \rho(\Phi_1, \Phi_1') = \int d\Phi_2 \;\Psi_+[\Phi_1,\Phi_2]\, \Psi_-[\Phi_1',\Phi_2] =  \frac{1}{Z(\beta)} \sum_n e^{-\beta E_n} \varphi_n\qty[\Phi_1]\varphi_n\qty[\Phi_1']\;.
\end{equation}
Here $\rho(\Phi_1,\Phi_1')$ should be understood as the representation of the density matrix on the surface $\mathscr{S}_1$, which denotes
a thermal state at temperature $\beta^{-1}$. We will comment more on this during our discussion of the TFD state for causal diamonds in the next section.\\

This completes the connection between the TFD state and the underlying geometry via the Euclidean path integral formalism. In the next section, we briefly describe the geometry of the causal diamond and appropriate coordinates for an observer constrained within this diamond (called the diamond observer) and then proceed to apply the formalism developed in this section to the causal diamond geometry.

\section{Thermofield Double State in causal diamonds}\label{sec3}
\subsection{Lorentzian metric of the causal diamond}\label{seccds}
\begin{figure}[t]
    \centering
    \begin{subfigure}{0.30\textwidth}
        \includegraphics[width=\linewidth]{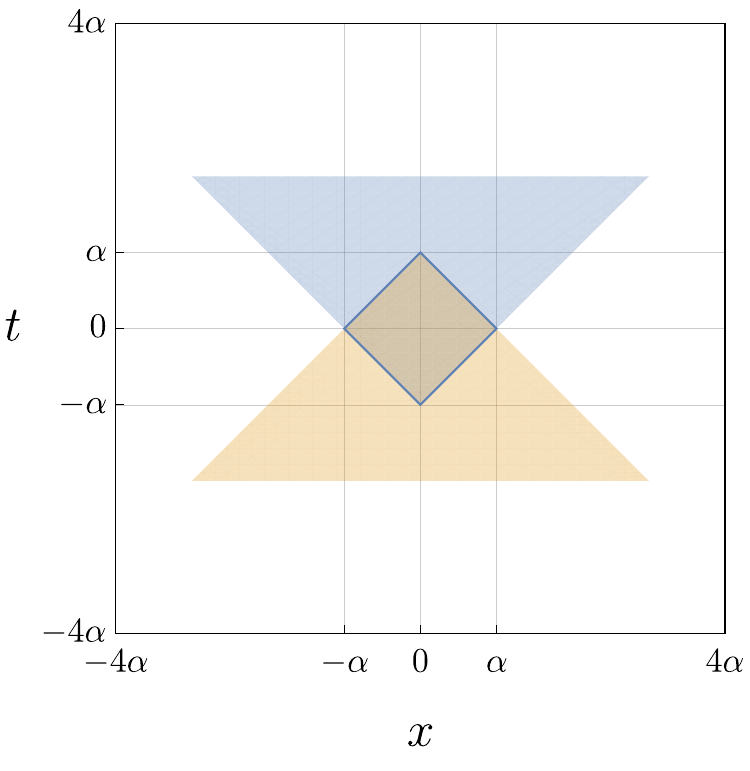}
\captionsetup{justification=centering} 
        \caption{}
        \label{fig:cd}
    \end{subfigure}
    \hfill
    \begin{subfigure}{0.3\textwidth}
        \includegraphics[width=\linewidth]{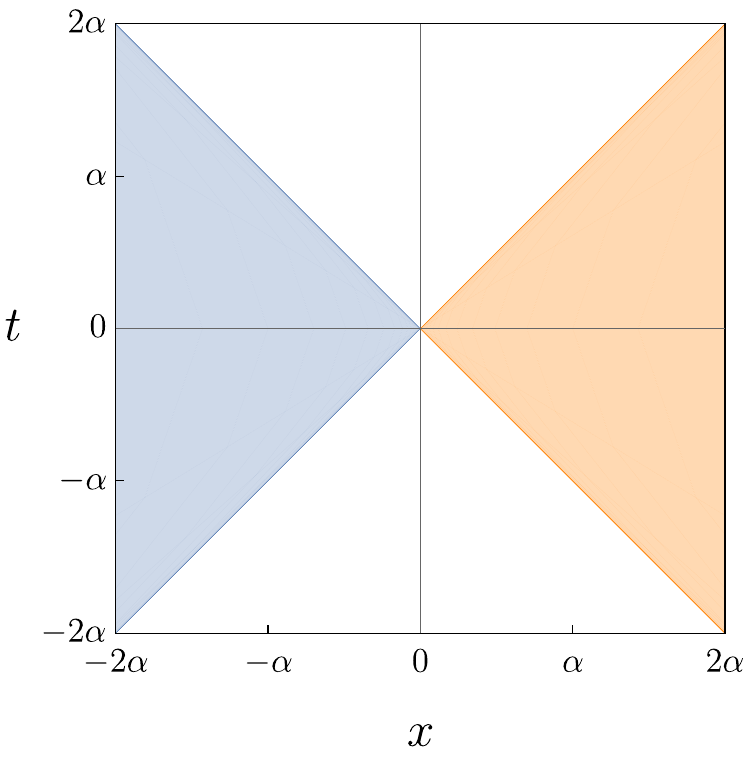}
\captionsetup{justification=centering}
        \caption{}
        \label{fig:rlr}
    \end{subfigure}
    \hfill
    \begin{subfigure}{0.3\textwidth}
        \includegraphics[width=\linewidth]{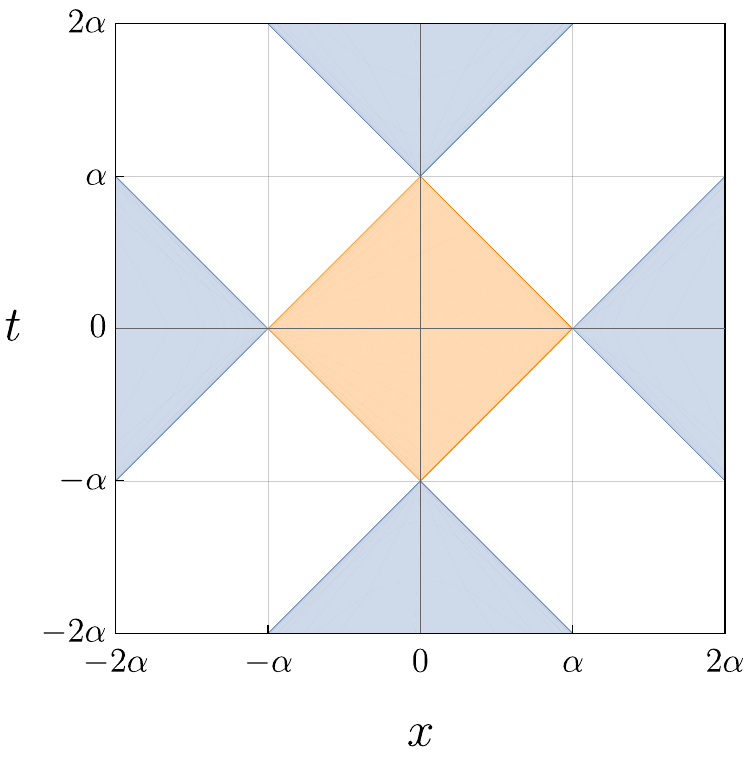}
\captionsetup{justification=centering}
        \caption{}
        \label{fig:cdie}
    \end{subfigure}    
    \caption{(a) Causal diamond for a lifetime of \(2\alpha\). (b) Right (orange) and left (blue) Rindler wedges. (c) Interior (orange) and exterior (blue) regions of the causal diamond.}
    \label{fig:diamond_rindler}
\end{figure}

The causal diamond is the spacetime region defined by the intersection of the future light cone corresponding to the birth event of an observer and the past light cone corresponding to the death event of the same observer at a future finite time (say, $2\alpha$) in the Minkowski spacetime. Therefore, we note that the diamond observer has a `lifetime' of \(2\alpha\) (see figure \ref{fig:cd}). As mentioned above, we will first define the diamond coordinates and its connection with the Rindler and Minkowski coordinates following \cite{Chakraborty:2022qdr} and \cite{camblong2024entanglement}.\\

To obtain the diamond coordinates, we note that the Minkowski coordinates restricted to the diamond region $(t_d, x_d)$ can be obtained from the Minkowski coordinates restricted to the right Rindler edge $(t_r,x_r)$ (see figure \ref{fig:diamond_rindler}) by using a special conformal transformation followed by a translation along the $-x$ axis. More specifically,
\begin{equation}
    (t_d,x_d)=T(-\alpha)\circ K\qty(\frac{1}{2\alpha})(t_r,x_r)\;.\label{comp}
\end{equation}
Here $K(\cdot)$ is the special conformal transformation and $T(\cdot)$ is the translation operation defined respectively as:
\begin{equation}
    K(a) x^\mu = \frac{x^\mu+ a\,\delta^\mu_1 (x\cdot x)}{1+2 a x^1 + a^2 (x\cdot x)}\quad,\quad T(a)\,x^\mu=(x^0,x^1+a)
    \label{s,.l;8},
\end{equation}
where $x\cdot x$ is the Minkowski squared norm of the 2-vector $x^\mu$ and \(\delta^\mu_1\) is the Kronecker delta, which is nonzero when $\mu = 1$. The special conformal transformation (SCT) given by \(K\qty(1/2\alpha)\) has the effect of compactifying the unbounded right Rindler wedge into a diamond of size \(2\alpha\) while preserving the causal structure. The spatial translation denoted by \(T(-\alpha)\) then shifts the center of the diamond to the origin by translating the spatial coordinates towards the $-x$ axis.\\

We note here that the left Rindler wedge (shown in blue in figure \ref{fig:rlr}), which is causally disconnected from the right Rindler wedge (shown in orange in figure \ref{fig:rlr}), can be mapped to the exterior region of the diamond (shown in blue in figure \ref{fig:cdie}) using a similar mapping to the conformal transformation \eqref{comp}. This region is causally disconnected to the interior of the diamond from the perspective of the observer constrained to the causal diamond. More details about the mapping from the left Rindler wedge to the exterior of the diamond can be found in \cite{camblong2024entanglement}.\\

As described in \cite{Chakraborty:2022qdr}, we now define the diamond coordinates $(\eta,\xi)$, which from the perspective of the diamond observer covers the entirety of the interior of the causal diamond region, i.e., $-\infty<\eta<\infty$ and $0<\xi<\infty$. These coordinates are inspired from the mapping between Rindler coordinates and Minkowski coordinates restricted to the right Rindler wedge as originally introduced by Unruh \cite{Unruh:1976db}. In terms of $(t_r,x_r)$, it is given by the relations:
\begin{equation}
    t_r=\xi\sinh(\frac{\eta}{\alpha})\quad,\quad x_r=\xi\cosh(\frac{\eta}{\alpha}),\label{cdtrrm}
\end{equation}
which can be recast in terms of $(t_d,x_d)$ using the mapping defined in \eqref{comp}. The explicit form as derived in \cite{Chakraborty:2022qdr} is given by:
\begin{equation}
    t_d=\frac{4 \alpha ^2 \xi  \sinh \left(\frac{\eta }{\alpha }\right)}{4 \alpha ^2+4 \alpha  \xi  \cosh \left(\frac{\eta }{\alpha }\right)+\xi ^2}\quad ,\quad x_d=\frac{\alpha  \xi ^2-4 \alpha ^3}{4 \alpha ^2+4 \alpha  \xi  \cosh \left(\frac{\eta }{\alpha }\right)+\xi ^2}.\label{tdxdtrxrtauxi}
\end{equation}
Using these transformations, we can rewrite the metric restricted to the causal diamond in terms of $(\eta,\xi)$ as
\begin{equation}
ds^2=\Omega^2(\eta,\xi)\qty[-\frac{\xi^2}{\alpha^2}d\eta^2+d\xi^2]\;.\label{len56}
\end{equation}
Here $\Omega^2(\eta,\xi)$ is a conformal factor dependent on the coordinates $(\eta,\xi)$. We note here that this form of the causal diamond metric is conformally equivalent to the Rindler metric which is given by the term in the bracket of \eqref{len56}. 

\subsection{Euclidean signature of the diamond metric}\label{eusign}
Since we need to calculate the Euclidean path integral, we now define the causal diamond metric using the Euclidean signature. 
By applying the Wick rotation \(\eta = -i\tau\), we can convert the Lorentzian diamond metric to the Euclidean metric:
\begin{equation}
    ds_E^2 = \Omega_E^2(\tau,\xi)\qty[\frac{\xi^2}{\alpha^2}d\tau^2+d\xi^2]\;, \label{eq:Euclid_line_cd}
\end{equation}
where the subscript $E$ denotes the Euclidean signature and the conformal factor is written with respect to the Euclidean time $\tau$
\begin{equation}
    \Omega_E^2(\tau,\xi) = \Omega^2(-i\tau,\xi) = \frac{16\alpha^4}{\qty(4\alpha\xi\cos\qty(\frac{\tau}{\alpha})+\xi^{2}+4\alpha^2)^2}\;.\label{fcon}
\end{equation}
Here, we again note that the Euclidean metric of the diamond is conformally equivalent to the Euclidean Rindler metric, which appears within the square brackets in \eqref{eq:Euclid_line_cd}.

\subsection{Exact solution of the Euclidean path integral}\label{3.2epi}
Let us now consider the Euclidean action of a real massless scalar field on a two-dimensional manifold,
\begin{equation}
    S_E\qty[\phi]=\frac{1}{2}\int\sqrt{g}d^2x g^{\mu\nu}\partial_\mu\phi\partial_\nu\phi.\label{euaci2}
\end{equation}

The Euclidean action \eqref{euaci2} is invariant under conformal transformations of the metric and the field (which in 2 dimensions has a conformal weight of 0). Specifically, for the transformations
\begin{equation}
    g^{\mu\nu} \to \Omega^{-2}g^{\mu\nu},\quad \sqrt{g} \to \Omega^{2}\sqrt{g},\quad \phi \to \phi,
\end{equation}
we obtain
\begin{equation}
    S_E\qty[\phi] \to \frac{1}{2}\int d^2x \, \Omega^{2} \sqrt{g} \, \Omega^{-2} g^{\mu\nu} \partial_{\mu}\phi \, \partial_{\nu}\phi = S_E\qty[\phi].\label{acinv}
\end{equation}
Therefore, we are dealing with conformal field theory (CFT).\\

Taking the geometry described by \(g_{\mu\nu}\) as that given by the line element in Eq.~(\ref{eq:Euclid_line_cd}), we find that the Euclidean action \eqref{euaci2} takes the following form:
\begin{equation}
    S_E\qty[\phi] = -\frac{1}{2}\int d\xi d\tau \phi\left[\frac{\alpha}{\xi}\partial_\tau^2+\frac{1}{\alpha}\partial_\xi+\frac{\xi}{\alpha}\partial_\xi^2\right]\phi.\label{euaci3}
\end{equation}

Let us now set \(\xi = \alpha e^{\rho/\alpha}\) in the Euclidean action \eqref{euaci3}, such that \(-\infty < \rho < \infty\), which gives us
\begin{equation}
    S_E\qty[\phi] = -\frac{1}{2}\int d\rho d\tau\phi\qty[\partial_\tau^2+\partial_\rho^2]\phi. \label{swphi}
\end{equation}

The action in Eq.~(\ref{swphi}) has the same form as it would if the field were defined on a Euclidean plane with coordinates \((\tau, \rho)\). This is due to the fact that the line element
\begin{align}
    ds_E^2 = \Omega_E^{\prime 2}(\tau,\rho) (d\tau^2 + d\rho^2)\nonumber\\
    \label{rindbueno33}
\end{align}
is conformal to the Euclidean plane, and the Euclidean action of the field is invariant under conformal transformations. The conformal factor in \eqref{rindbueno33} is given by
\begin{equation}
    \Omega_E^{\prime 2}(\tau,\rho)=\frac{16 e^{\frac{2 \rho }{\alpha }}}{\left(4 e^{\rho /\alpha } \cos \left(\frac{\tau }{\alpha }\right)+e^{\frac{2 \rho }{\alpha }}+4\right)^2},
\end{equation}
and it relates to the conformal factor in \eqref{fcon} via \(\Omega_E^{\prime 2}(\tau,\rho)=\Omega_E^{2}(\tau,\alpha e^{\rho/\alpha})e^{\frac{2 \rho }{\alpha }}\).\\

It is important to mention that a fundamental requirement of the Laflamme method is the periodicity \(\beta\) in the Euclidean time coordinate, which provides the cylindrical structure of the manifold. Therefore, since the action \eqref{swphi} can be interpreted as if it were on the background geometry of a Euclidean plane given by \(\qty{\tau,\rho}\), imposing the periodicity in \(\tau\) given by \(\tau \sim \tau + \beta\) precisely yields the Laflamme cylinder (figures \ref{fig31}, \ref{fig32}, and \ref{fig33}). Now that we have identified the coordinate \(\tau\) for the Laflamme cylinder in the Euclidean action \eqref{swphi}, we can divide this action over the submanifolds \(M_\pm\). As we know, this is possible by applying two cuts in \(\tau\), giving us the surfaces \(\mathscr{S}_1 = \tau_1\) and \(\mathscr{S}_2 = \tau_2\), such that \(\tau_2 = \tau_1 + \beta/2\). Thus, for \(M_+\), we have that in the Euclidean action, the temporal coordinate \(\tau\) will take values in the range \(0 < \tau < \beta/2\), while for \(M_-\), we have \(\beta/2 < \tau < \beta\).\\

Similarly, since the scalar field is free, we can bring the action into the desired form by applying the Fourier transform to the coordinate $\rho$ of the scalar field:
\begin{equation}
\phi(\tau,\rho)=\frac{1}{\sqrt{2\pi}}\int e^{i\rho\lambda}\Tilde{\phi}(\tau,\lambda)d\lambda.
\end{equation}
Thus, by substituting $\phi(\tau,\rho)$ into the action \eqref{swphi}, we obtain
\begin{equation}
S_E\qty[\phi]=-\frac{1}{2}\int d\tau d\lambda\Tilde{\phi}(\tau,\lambda)\qty[\partial_\tau^2-\lambda^2]\Tilde{\phi}^*(\tau,\lambda).\label{sjl2d9}
\end{equation}

Since the field $\phi(\tau,\rho)$ is real, the complex field in the Fourier transform obeys the condition $\Tilde{\phi}^*(\tau,\lambda)=\Tilde{\phi}(\tau,-\lambda)$. From this condition, we derive the parity conditions in $\lambda$ for the real and imaginary components of the field:
\begin{equation}
    \Re[\Tilde{\phi}(\tau, -\lambda)]=\Re[\Tilde{\phi}(\tau, \lambda)]\quad,\quad \Im[\Tilde{\phi}(\tau, -\lambda)]=-\Im[\Tilde{\phi}(\tau, \lambda)].
\end{equation}

Thus, we note that it is possible to rewrite the action \eqref{sjl2d9} for the real scalar field $\psi(\tau, \lambda)$ (see \ref{apemn1} for the details of the computation), defined as
\begin{equation}
    \psi(\tau, \lambda) := \Tilde{\phi}_r(\tau, \lambda) \sqrt{1 + f(\lambda)^2},
\end{equation}
such that the real and imaginary parts of $\Tilde{\phi}(\tau, \lambda)$ satisfy
\begin{equation}
    \Re[\Tilde{\phi}(\tau, \lambda)]=\Tilde{\phi}_r(\tau, \lambda)\quad,\quad \Im[\Tilde{\phi}(\tau, \lambda)]=f(\lambda)\Tilde{\phi}_r(\tau, \lambda),
\end{equation}
where $f(\lambda)$ is a real odd function in $\lambda$.\\

In this way, the action \eqref{sjl2d9} can be written as
\begin{equation}
    S_E[\psi] = -\frac{1}{2} \int d\tau d\lambda \, \psi(\tau, \lambda) \left[\partial_\tau^2 - \lambda^2 \right] \psi(\tau, \lambda).
\end{equation}

Then, following the method described by Laflamme \cite{Laflamme:1988wg,Laflamme:1989cm}, we discretize the path integral for each mode \({\lambda_\alpha}\) such that the Euclidean action on each submanifold \(M_\pm\) corresponds to that of a harmonic oscillator:
\begin{equation}
    S_E\qty[\psi_{\lambda_\alpha}] = -\frac{1}{2}\int_{\mathscr{C}_\pm} d\tau \psi_{\lambda_\alpha}(\tau)\qty[\partial_\tau^2 -{\lambda_\alpha}^2]\psi_{\lambda_\alpha}(\tau), \label{final21}
\end{equation}
where \(\psi_{\lambda_\alpha}(\tau) = \psi(\tau, \lambda_\alpha)\sqrt{\Delta\lambda} = \Tilde{\phi}_r(\tau, \lambda_\alpha) \sqrt{1 + f(\lambda_\alpha)^2}\sqrt{\Delta\lambda}\), and notice that the interval $\Delta \lambda = \lambda_{\alpha+1} - \lambda_\alpha> 0$.\\

We now look back at the quantity we want to calculate, the wave functionals $\Psi_{\pm}$ defined in \eqref{psi+++} and \eqref{psi---}. Since the action for each frequency mode $\lambda_\alpha$ only depends on $\lambda_\alpha$, and is independent of other frequency modes, we only consider the action $S_E\qty[\psi_{\lambda_\alpha}]$ \cite{Laflamme:1988wg,Laflamme:1989cm}, instead of the full action $S_E[\phi]$. Since we will consider a single frequency, we will drop the index $\alpha$ from the frequency hereafter. In order to be consistent, we now redefine our wave functionals for $\psi_\lambda$ instead of the full field configuration:
\begin{equation}
   \Psi_\pm[\Phi_1,\Phi_2] := \Psi_\pm[\psi_1,\psi_2] = \int_{\mathscr{C}_{\pm}} D\psi_\lambda\; e^{-S_E[\psi_\lambda]}
\end{equation}
Here $\psi_{1,2}$ are the boundary configurations of the field $\psi_\lambda$ specified on the surfaces \(\mathscr{S}_1=\tau_1\) and \(\mathscr{S}_2=\tau_2\), respectively.\\

Due to the quadratic nature of the field in the Euclidean action \eqref{final21}, the path integral \(\Psi_\pm\qty[\psi_1,\psi_2]\) for the action \(S_E\qty[\psi_{\lambda}]\) of the regular fields \(\psi_{\lambda}\) on the submanifold \(M_\pm\) with boundary conditions \(\psi_1\) and \(\psi_2\) , can be computed exactly (see, for example, \cite{feynman2010quantum,kleinert2006path}). In this way, for \(\Psi_+\qty[\psi_1,\psi_2]\) where \(\Delta\tau=\tau_2-\tau_1=\beta/2\) (the treatment is similar for \(\Psi_-\qty[\psi_1,\psi_2]\), where \(\Delta\tau=\tau_1-\tau_2=(\tau_1+\beta)-(\tau_1+\beta/2)=\beta/2\)), we have
\begin{equation}
\Psi_+\qty[\psi_1,\psi_2]=\qty(\frac{\lambda}{2\pi\sinh\qty(\frac{\lambda\beta}{2})})^{1/2}\exp\qty[-\frac{\lambda}{2}\qty(\left(\psi_1^2+\psi_2^2\right)\coth\qty(\frac{\lambda\beta}{2})-\frac{2\psi_1\psi_2}{\sinh\qty(\frac{\lambda\beta}{2})})],\label{k+solu}
\end{equation}

We note here that the expression given in \eqref{k+solu} is well known in the literature due to it being the propagator for a harmonic oscillator. However, the main point of novelty introduced in this section is the casting of the causal diamond metric into the conformal form \eqref{rindbueno33} such that we can write the Euclidean action for the causal diamond in a quadratic form as given in \eqref{final21}, so that we can use the harmonic oscillator result.\\

We observe that by applying the following property of Hermite polynomials of \( n \)-th order
\begin{equation}
\frac{\exp\qty[-\frac{\xi^2+\eta^2-2\xi\eta\zeta}{(1-\zeta^2)}]}{\sqrt{1-\zeta^2}}=e^{-(\xi^2+\eta^2)}\sum_{n=0}^\infty \frac{\zeta^n}{2^n n!}H_n(\xi)H_n(\eta),
\end{equation}
with $\xi\to\sqrt{\lambda}\psi_1$, $\eta\to\sqrt{\lambda}\psi_2$ and $\zeta\to e^{-\beta\lambda/2}$, we can reformulate the expression given in \eqref{k+solu} in terms of the \( n \)-excited state of the harmonic oscillator applied to the boundary values of the field,
\begin{equation}
     \varphi_n[\psi_{1,2}] = \bra{\psi_{1,2}}\ket{n}=\qty(\sqrt{\frac{\lambda}{\pi}}\frac{1}{2^n n!})^{1/2}H_n(\sqrt{\lambda}\psi_{1,2})e^{-\frac{\lambda}{2}\psi_{1,2}^2},
\end{equation}
which is a wave functional of those values and is expressed in terms of the Hermite polynomial of \( n \)-th order \( H_n(\sqrt{\lambda}\psi_{1,2}) \). Thus, the normalized form of \(\Psi_+\qty[\psi_1,\psi_2]\) can be written as
\begin{equation}
    \Psi_+\qty[\psi_1,\psi_2]=\frac{1}{\sqrt{Z\qty(\beta)}}\sum_{n=0}^{\infty}e^{-\frac{\beta}{2}E_n}\varphi_n\qty[\psi_1]\varphi_n\qty[\psi_2],\label{tfd12}
\end{equation}
where \( E_n = \lambda \left(n + 1/2 \right) \) is the eigenvalue of the Hamiltonian operator \( H \), and \( Z(\beta) \) is the normalization factor that depends on \(\beta\) and is given by
\begin{equation}
    1=\int_{-\infty}^{\infty}d\psi_1d\psi_2\Psi_+\qty[\psi_1,\psi_2]\Psi^*_+\qty[\psi_1,\psi_2]=\frac{1}{Z}\sum_{n=0}^{\infty}e^{-\beta E_n}\Rightarrow{} Z=\sum_{n=0}^{\infty}e^{-\beta E_n},
\end{equation}
where the orthogonality condition for Hermite polynomials, given by
\begin{equation}
    \int H_n(\sqrt{\lambda}\psi) H_{n'}(\sqrt{\lambda}\psi) e^{-\lambda \psi^2} \, d\psi = \sqrt{\frac{\pi}{\lambda}} \, 2^n \, n! \, \delta_{n,n'},
\end{equation}
has been used.\\

At this point, we note that the Euclidean path integral in \eqref{tfd12} corresponds to the wave functional for the vacuum state introduced by Umezawa and Takahashi, denoted as $\ket{0(\beta)}$, and better
known as the thermofield double state (TFD). The meaning of this expression
will be further explained in the next section.

\subsection{Diamond temperature for static observer}\label{temp159--}

From \eqref{tfd12}, we already see that mapping the causal diamond to the cylindrical geometry with periodicity $\beta$ in the Euclidean time yields a TFD state with temperature $\beta^{-1}$. However, so far, we have not determined what this periodicity $\beta$ should be. To identify $\beta$, we note that the two disconnected surfaces $\mathscr{S}_1$ and $\mathscr{S}_2$ that we use to split open the cylinder, corresponds to the birth and death event of the diamond observer, because the presence of the two events create the two causally disconnected regions (shown in blue and orange in figure \ref{fig:cdie}). Hence the two edges of the unrolled cylinder $\tau = -\beta/2$ and $\tau = + \beta/2$ corresponds to $t_E = -\alpha$ and $t_E = \alpha$ in the Euclidean plane. Here, $t_E$ denotes the Euclideanized version of $t_d$. We now use the relation between $t_E$ and $\tau$ to identify $\beta$:
\begin{equation}
    \tau = \alpha \tan^{-1}\left( \frac{4\alpha^2 t_{E}}{4\alpha^2(x_{\!_d}+\alpha) - 2\alpha[(x_{\!_d}+\alpha)^2 + t_{E}^2]}\right)\;.
\end{equation}
This relation can be obtained from the definition of the coordinate $\tau$ and the mapping used to relate $(t_d,x_d)$ with $(\eta,\xi)$, and is shown in detail in \ref{t-m.,k}. From this relation, for a static observer at $x=0$, which we will call the diamond observer, we can write:
\begin{equation}
    \tau(t_E = -\alpha, x =0) = -\pi \alpha/2 \quad,\quad \tau(t_E = \alpha, x =0) = \pi \alpha/2.
\end{equation}
Identifying these edges with $-\beta/2$ and $\beta/2$ respectively, we then find the diamond temperature for a static observer at $x=0$ as 
\begin{equation}
    \beta = T^{-1} = \pi\alpha\;.
\end{equation}
This is consistent with what has been observed in the literature\cite{Martinetti:2002sz, Chakraborty:2022qdr,Su:2015oys}.\\

As a side note, we also want to mention here that the periodicity $\beta$ can also be argued from the non-existence of conical singularity and identifying the Euclidean time coordinate with the angular coordinate, as is usually done for Hawking temperature derivation \cite{hartman2015lectures}. However, due to the presence of the conformal factor, it becomes a bit tricky. To avoid any confusion, we identify the edges of the disconnected surfaces with the edges of the causal diamond to find $\beta$.\\

In the next section, we briefly describe the meaning of the TFD state for a causal diamond before concluding this article.

\section{Analysis of the TFD state and the density matrix for the causal diamond}\label{tfdss}

\subsection{The TFD state for the causal diamond}
To identify the states defined by the field configurations on the boundary, \(\psi_{1,2}\), within the context of the causal diamond, we first need to analyze the Laflamme method in Rindler spacetime, which is detailed in \ref{anlysnrin}. This approach is necessary because the metrics of the causal diamond and Rindler spacetime are related through the conformal transformation \eqref{comp}, enabling us to implement this transformation at the quantum level by leveraging the conformal invariance of the action.\\

Since we are dealing with a CFT, the generators of special conformal transformations and translations used in \eqref{comp} are symmetries of the action. Consequently, the vacuum state remains invariant under conformal transformations, which are implemented by the following unitary operator \cite{DiFrancesco:1997nk}:
\begin{equation}
    U = e^{-ia^\mu \Tilde{P}_\mu}e^{-ib^\mu \Tilde{K}_\mu},\label{uoperu729}
\end{equation}
where \(\Tilde{P}\) and \(\Tilde{K}\) are the generators of translations and special conformal transformations in the Hilbert space of quantum states, with transformation parameters \(a^\mu = (0, -\alpha)\) and \(b^\mu = \left(0, -1/2\alpha\right)\).\\

Since the Rindler and causal diamond geometries are conformally related, we have the following transformations for the fields, $\psi$, and states, $\ket{\psi}$, between Rindler spacetime, in the right (R) and left (L) wedges, and the interior (in) and exterior (ext) regions of the causal diamond, respectively:
\begin{equation}
    U\psi_{R, L}U^{\dagger} = \psi_{in,\,ext}, \quad U\ket{\psi_{R, L}} = \ket{\psi_{in,\,ext}}.\label{uprlu-1inext}
\end{equation}

From the analysis in Rindler, we have $\Tilde{\psi}_R = \Theta\psi_L\Theta^{-1}$ (see equation \eqref{mimeq616}), where \(\Theta\) is the anti-unitary \(CPT\) operator, which implements a natural map between the Hilbert spaces for the fields on the left \((x<0)\) \(\mathcal{H}_L\) and on the right \((x>0)\) \(\mathcal{H}_R\) \cite{Harlow:2014yka}. Therefore, from \eqref{mimeq616} and \eqref{uprlu-1inext} we observe that  
\begin{equation}
    (U\Theta U^\dagger)\psi_{ext}(U\Theta U^\dagger)^{-1} = \Tilde{\psi}_{in}.
\end{equation}
Therefore, the operator $U\Theta U^\dagger$ provides a natural map between the Hilbert spaces for the fields in the interior region of the causal diamond, \(\mathcal{H}_{in}\), and those in the exterior region, \(\mathcal{H}_{ext}\). Thus, we define
\begin{equation}
    \Theta_D:=U\Theta U^\dagger \quad\Rightarrow\quad \Tilde{\psi}_{in}=\Theta_D\psi_{ext}\Theta_D^{-1}\quad,\quad \ket*{\Tilde{\psi}_{in}}=\Theta_D\ket{\psi_{ext}}\label{thetaddefi}.
\end{equation}

Moreover, the transition amplitude for the Rindler case is given by $\bra{\psi_L,\psi_R}\ket{0(\beta)} \propto \bra{\psi_R}e^{-\frac{\beta_r}{2} H_r}\Theta\ket{\psi_L}$ (see equation \eqref{sdaesd41}), therefore, applying \eqref{thetaddefi} and \eqref{uprlu-1inext} to this expression, and using the fact that \(\pi H^r_\theta = (\beta_r/2) H_r\) (where \( H_r \) is the Hamiltonian of the field in the right wedge of Rindler, \( H_\theta \) generates rotations in the Euclidean plane, and \( \beta_r \) is the periodicity of the Euclidean time coordinate in Rindler), we have
\begin{align}
\bra{\psi_L,\psi_R}\ket{0(\beta)} &\propto \bra{\psi_R}e^{-\frac{\beta_r}{2} H_r}\Theta\ket{\psi_L}\nonumber\\
    &=(\bra{\psi_R}U^\dagger)(U e^{-\frac{\beta_r}{2} H_r}U^\dagger)( U\Theta U^\dagger)(U\ket{\psi_L})\nonumber\\
    &=\bra{\psi_{in}}(U e^{-\frac{\beta_r}{2} H_r}U^\dagger)\Theta_D\ket{\psi_{ext}},\label{2-f.;}
\end{align}
where $(U e^{-\frac{\beta_r}{2}H_r}U^\dagger)$ is the conformal transformation of the angular evolution operator for Rindler states from the left $(x<0)$ to the right $(x>0)$ of the Euclidean plane. This transformation, as inferred from \eqref{2-f.;}, will provide the operator responsible for angularly propagating states from the exterior region to the interior of the causal diamond. Therefore, this transformation will be of the form
\begin{equation}
    e^{-\frac{\beta}{2}H}=U e^{-\frac{\beta_r}{2}H_r}U^\dagger,
\end{equation}
where $\beta$ is the periodicity for the $\tau$ coordinate of the causal diamond. Thus, the transition amplitude will be given by
\begin{equation}
    \bra{\psi_{in}}e^{-\frac{\beta}{2}H}\Theta_D\ket{\psi_{ext}}.\label{ampltuhcd}
\end{equation}

We know that \( H \) in \eqref{ampltuhcd} is responsible for implementing at the quantum level the geometric flow given by the vector field \(\partial_\tau\) (see, for example, \cite{Jacobson:2015hqa}), which generates evolution in \(\tau\) in the Euclidean causal diamond. Its explicit form can be obtained by applying the Wick rotation \(\eta = -i\tau\) to the coordinate transformation between the causal diamond and Minkowski spacetime \eqref{tdxdtrxrtauxi}:
\begin{equation}
    \partial_{\tau}=\frac{1}{\alpha^2}\qty(t_E x\partial_x + \frac{1}{2}(\alpha^2+t_E^2-x^2)\partial_{t_E}).\label{ccwea}
\end{equation}

\begin{figure}[t]
    \centering
    \begin{subfigure}{0.2888\textwidth}
        \includegraphics[width=\linewidth]{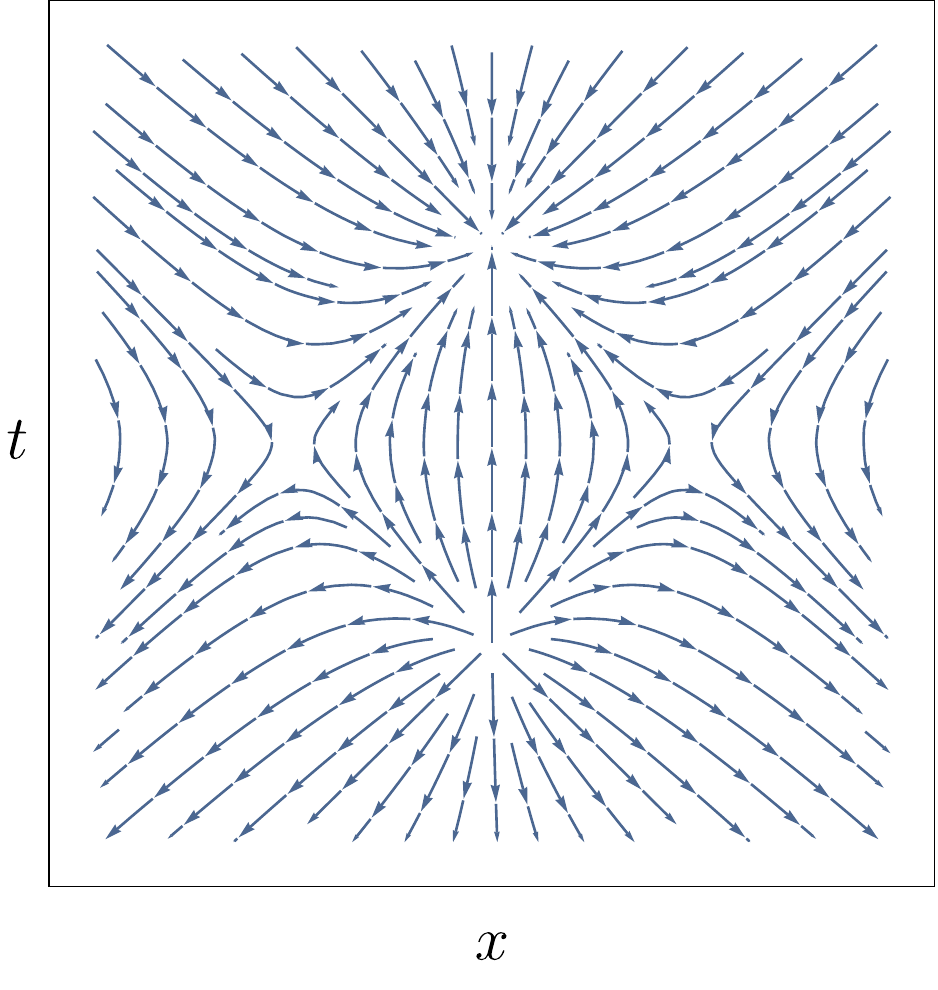}
\captionsetup{justification=centering} 
        \caption{}
        \label{fig8a}
    \end{subfigure}
    \hfill
    \begin{subfigure}{0.3\textwidth}
        \includegraphics[width=\linewidth]{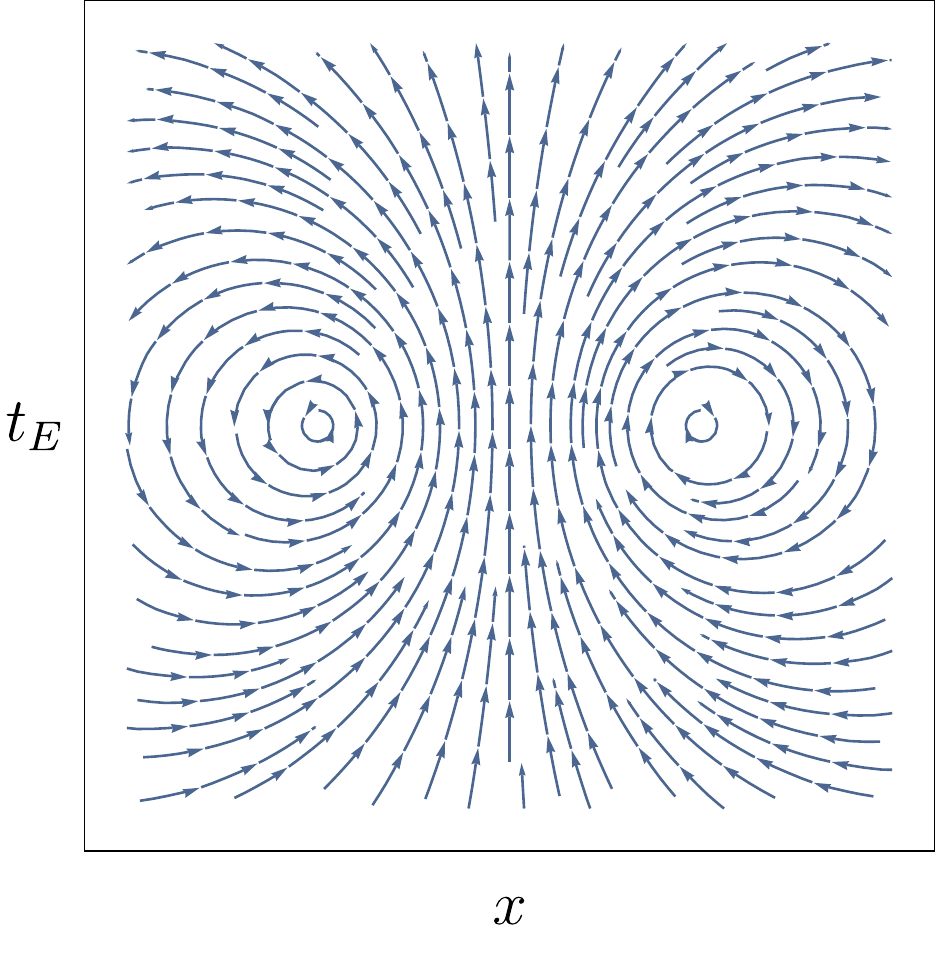}
\captionsetup{justification=centering}
        \caption{}
        \label{fig8b}
    \end{subfigure}
    \hfill
    \begin{subfigure}{0.3\textwidth}
        \includegraphics[width=\linewidth]{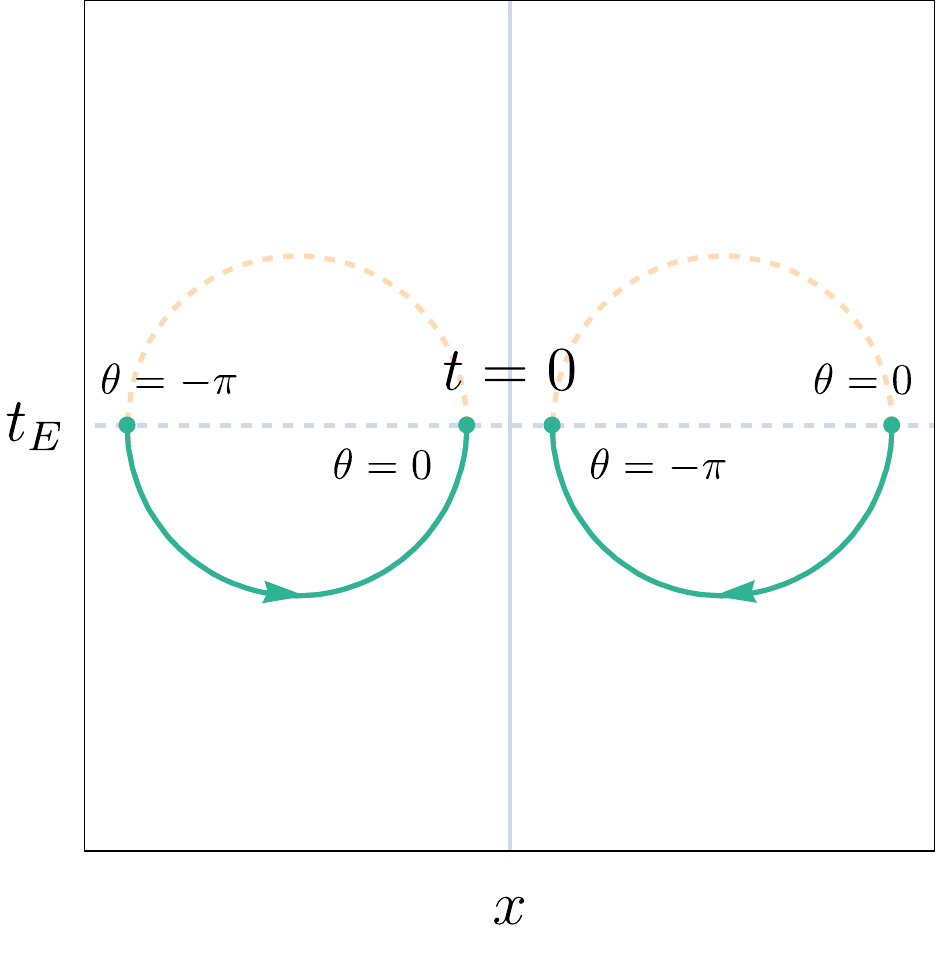}
\captionsetup{justification=centering}
        \caption{}
        \label{fig8c}
    \end{subfigure}    
    \caption{Integral curves of the vector field (a) $\partial_\eta$ in Minkowski, and (b) $\partial_\tau$ in the Euclidean Plane. (c) Evolution by $\pi$ radians in the lower half of the Euclidean plane.}
    \label{}
\end{figure}

It is important to mention that the vector field \(\partial_\tau\) is a conformal Killing vector of the Euclidean causal diamond metric \cite{Arzano:2020thh,Arzano:2021cjm,Arzano:2023pnf,arzano2024entanglement,camblong2024entanglement}, which is related to the Killing vector in Rindler, responsible for generating evolution in the Euclidean temporal coordinate, through the conformal transformation \eqref{comp}. The integral curves of \(\partial_\tau\) are shown in figure \ref{fig8b}. Similarly, in Lorentzian signature, given by \(\partial_\eta\), we have that it generates transformations of the diamond onto itself, as shown in figure \ref{fig8a} (in \cite{Martinetti:2002sz}, the method used by the authors implements the modular operator responsible for the modular flow associated with the mentioned transformation).\\

As we can notice, the vector field \(\partial_\tau\) implements rotations from the region \(\abs{x} > \alpha\) to the region \(\abs{x} < \alpha\) in the lower part of the Euclidean plane. Specifically, we observe two rotations: one in the region $x<0$ in the counterclockwise direction and another in the region $x>0$ in the clockwise direction (see figure \ref{fig8c}). The propagation directions are consistent with the transition amplitude \eqref{ampltuhcd}, since the Hamiltonian will generate the propagation of states from the exterior region to the interior of the causal diamond at $t=t_E=0$.\\

Returning to equation \eqref{2-f.;}, let us analyze the left side of this expression by inserting the operator \(1 = (U \otimes U)^\dagger (U \otimes U)\):
\begin{equation}
    \bra{\psi_L,\psi_R}\ket{0(\beta)}=\bra{\psi_L,\psi_R}(U\otimes U)^\dagger(U\otimes U)\ket{0(\beta)}.
\end{equation}

Recall that the states in the above expression are defined in the Hilbert space given by $\mathcal{H}_{ext}\otimes\mathcal{H}_{in}$. Furthermore, since we are working in a conformal field theory, the Minkowski vacuum state is invariant under the action of $(U \otimes U)$, while the state $\ket{\psi_L, \psi_R}$ is affected by $U \otimes U$ as specified in \eqref{uprlu-1inext}. Therefore, we have
\begin{equation}
    \Psi_+[\psi_{ext},\psi_{in}]=\bra{\psi_{in}}e^{-\frac{\beta}{2}H}\Theta_D\ket{\psi_{ext}}\propto\bra{\psi_{ext},\psi_{in}}\ket{0(\beta)}.
\end{equation}

Finally, proceeding similarly to the Rindler analysis, inserting the operator $1=\sum_n\ket{n_{in}}\bra{n_{in}}$, where $\ket{n_{in}}$ are the eigenbasis of the Hamiltonian in the interior region of the causal diamond, we obtain the following normalized state:
\begin{equation}
    \ket{0(\beta)} = \frac{1}{\sqrt{Z\qty(\beta)}} \sum_{n=0}^{\infty} e^{-\frac{\beta}{2}E_n} \ket{n_{in}} \otimes \Theta_D^{-1}\ket{n_{in}}.\label{tfdsrej22}
\end{equation}

We can observe that this expression allows us to fully identify the result obtained in \eqref{tfd12} through $\Psi_+[\psi_{ext},\psi_{in}] = \bra{\psi_{ext},\psi_{in}}\ket{0(\beta)}$, where the boundary values are given by \(\psi_1 = \Tilde{\psi}_{in} = \Theta_D \psi_{ext} \Theta_D^{-1}\) and \(\psi_2 = \psi_{in}\). Moreover, the expression \eqref{tfdsrej22} represents the thermofield double state. This enables us to describe the thermal nature of the Minkowski ``vacuum" state as a pure state in terms of the direct product of the eigenbasis of the Hamiltonian for the field in the interior, $\ket{n_{in}}$, and exterior, $\ket{\Tilde{n}_{ext}} = \Theta_D^{-1} \ket{n_{in}}$, regions of the causal diamond.

\subsection{The density matrix for the physical system}\label{secdenma}
The TFD state \eqref{tfdsrej22} is described in terms of the Hamiltonian eigenbasis in the interior and exterior regions. This is because it corresponds to the purification of the mixed state describing the physical system. In this context, the state $\ket{\Tilde{n}_{ext}}$ corresponds to the fictitious field introduced by Umezawa and Takahashi. In this analysis, the fictitious states arise naturally when implementing considerations related to Euclidean geometry. Recall that to divide the manifold \( M \) of the cylinder, two cuts are made on the surfaces \(\{\mathscr{S}_1, \mathscr{S}_2\}\), and the field configuration is defined on these surfaces.\\

Since the density matrix for the pure state given by the TFD state \eqref{tfdsrej22} is \(\rho = \ket{0(\beta)}\bra{0(\beta)}\), the density matrix for the physical system \(\psi_{in}\) is given by
\begin{align}
    \rho_{in}&=\Tr_{ext}\rho=\sum_{m}\bra{\Tilde{m}_{ext}}\rho\ket{\Tilde{m}_{ext}}=\sum_{m}\bra{\Tilde{m}_{ext}}\ket{0(\beta)}\bra{0(\beta)}\ket{\Tilde{m}_{ext}}\nonumber\\
    &=\frac{1}{Z(\beta)}\sum_{m,n,n'}e^{-\frac{\beta}{2}(E_n+E_{n'})}\ket{n_{in}}\bra{\Tilde{m}_{ext}}\ket{\Tilde{n}_{ext}}\bra{n'_{in}}\bra{\Tilde{n}'_{ext}}\ket{\Tilde{m}_{ext}}\nonumber\\
    &=\frac{1}{Z(\beta)}\sum_n e^{-\beta E_n}\ket{n_{in}}\bra{n_{in}}=\frac{e^{-\beta H}}{Z(\beta)}.
\end{align}

Therefore, the vacuum state perceived by the observer confined to the interior region of the causal diamond is expressed through the density matrix \(\rho_{in}\), which describes a Gibbs state, i.e., a mixed ensemble in thermal equilibrium with a heat bath at a temperature given by \(\beta^{-1} = 1/\pi\alpha\). Additionally, the reduced state is thermal with respect to the notion of time translation defined by the flow of the conformal Killing vector field \(\partial_\eta\), which is timelike and future-oriented inside the causal diamond. This geometric flow is implemented at the quantum state level through the Hamiltonian \(H\). Finally, since \(\rho_{in}\) describes a mixed ensemble, the TFD state is an entangled state of the Hamiltonian eigenbasis \(\{\ket{n_{in}}, \ket{\tilde{n}_{ext}} = \Theta_D^{-1} \ket{n_{in}}\}\).

\section{Conclusions}\label{concl}This article contributes to a series of studies on the temperature of causal diamonds. The temperature perceived by a diamond observer has been derived using various approaches, including the modular flow and thermal time hypothesis \cite{Martinetti:2002sz}, Bogolyubov transformations \cite{Su:2015oys}, the Unruh-deWitt detector formalism \cite{Su:2015oys,Foo:2020rsy}, and the open quantum systems approach \cite{Chakraborty:2022qdr}, among others. In this paper, a path integral formalism is employed. All these derivations yield the same diamond temperature, confirming the robustness of this result. This work achieves two goals: demonstrating and popularizing the elegance and universality of Laflamme's framework for describing thermofield double (TFD) state and confirming the emergence of thermal properties akin to the Unruh effect without requiring accelerated observers. This underscores the need for further investigations into quantum effects in causal diamonds.\\

It is worth mentioning that it has recently been established that conformal Killing vectors in the causal diamond geometry are closely related to the generators of the $sl(2,\mathbb{R})$ algebra in (0+1)-D conformal field theory, known as conformal quantum mechanics \cite{Arzano:2020thh, Arzano:2021cjm, Arzano:2023pnf, DeLorenzo:2017tgx, DeLorenzo:2018ghq, herrero1999radial, Jacobson:2015hqa, arzano2024entanglement}. Moreover, conformal quantum mechanics plays a crucial role in determining the temperature of black hole radiation \cite{Camblong:2004ye, Camblong:2004ec, Camblong:2020pme, Azizi:2020gff, Azizi:2021qcu, Azizi:2021yto}, thereby linking causal diamonds more closely with the near-horizon physics of black holes. Causal diamonds are also intriguing due to their association with entanglement entropy in many-body systems and quantum chaos. These connections suggest that the understanding of thermal aspects of causal diamonds may be important in understanding the origins of black hole entropy and information scrambling. Given the importance of conformal Killing vectors in the present work (equation \eqref{ccwea}), it would be interesting to explore any potential connections between the approach and results of this article and these phenomena. We intend to explore some of these topics in the future.

\ack
Two of the authors, C.R.O. and G.V-M. would like to thank Pablo Lopez-Duque for illuminating discussions on the geometry of causal diamonds. C.R.O. and G.V-M. were partially supported by the Army Research Office (ARO), Grant W911NF-23-1-0202.

\appendix
\section{Temperature corresponding to the horizontal cut of the hypercylinder}\label{app:horizontal_cut_temp}
The surface \(\mathscr{S} = x_0 = 0\) shown in figure \ref{fig20} (there is nothing special about the value \(x_0=0\); any other finite value would lead to the same conclusion) provides us with two cylinders similar to the original one. Moreover, by identifying \(x\) and \(\tau\) with the Lorentzian metric \(ds^2 = -d\sigma^2 + d\tau^2\) through a Wick rotation \(x = i\sigma\) 
, we find that \(x\) is the Euclidean equivalent of the Lorentzian time \(\sigma\). Therefore, \(x\) will be referred to as ``Euclidean time." Consequently, the Euclidean equivalent of the time evolution operator for Lorentzian time \(\sigma\) is given by the operator \(U(\Delta x)=e^{-\Delta x H}\), which will be responsible for implementing the evolution in \(x\) on the Euclidean plane, that is, the evolution in Euclidean time. Thus, \(\Psi_+\) is the transition amplitude generated by \(U(\Delta x)\) for field configurations on the surfaces \(x = 0\) and \(x \to \infty\) (see the left cylinder in figure \ref{fig21}), while \(\Psi_-\) corresponds to those on the surfaces \(x \to -\infty\) and \(x = 0\) (see the right cylinder in figure \ref{fig21}). For both transition amplitudes, the vacuum state will represent the primary contribution, as the amplitude can be expressed as a summation over \(n\) that decays exponentially due to the factor \(e^{-E_n \Delta x}\), where \(E_n\), which increases with \(n\), are the eigenvalues of the Hamiltonian operator \(H\). Therefore, for \(\Delta x \to \infty\), the terms with \(n > 0\) will have a negligible contribution. For example, if for \(\Psi_+\) we have \(\Phi(x=0) = \phi_1\) and \(\Phi(x \to \infty) = \phi_2\), then
\begin{align}
    \Psi_+[\phi_1, \phi_2] &= \langle \phi_2 | e^{-\Delta xH} | \phi_1 \rangle = \sum_{n,n'} \langle \phi_2 | n' \rangle \bra{n'} e^{-\Delta xH} \ket{n}\bra{n}\ket{\phi_1}\nonumber\\
    &=\sum_n \varphi_{n}\qty[\phi_2]\varphi^*_{n}\qty[\phi_1]e^{-\Delta xE_n}\approx \varphi_{0}\qty[\phi_2]\varphi^*_{0}\qty[\phi_1] e^{-\Delta xE_0 },\label{gst01}
\end{align}
where we have expressed the states \(\ket{\phi_{1,2}}\) in the eigenbasis of the Hamiltonian operator, such that \(\varphi_{0}\qty[\phi_{1,2}] = \langle \phi_{1,2} | 0 \rangle\), and \(E_0\) is the vacuum state energy eigenvalue. In section \ref{tfdss}, we will show that the temperature associated with the thermofield double structure of the vacuum state is given by the inverse of the Euclidean time interval. Therefore, since in the present configuration the time interval given by \(\Delta x\) is infinite in \eqref{gst01}, there is no associated temperature: \(T = 0\).

\section{Euclidean action and path integral}\label{apemn1}
Using the Fourier transformation for the real scalar field $\phi$,
\begin{equation}
\phi(\tau,\rho)=\frac{1}{\sqrt{2\pi}}\int e^{i\rho\lambda}\Tilde{\phi}(\tau,\lambda)d\lambda,\label{tfptr}
\end{equation}
in the action \eqref{swphi}, we have
\begin{align}
S_E\qty[\phi] &= -\frac{1}{2}\int d\rho d\tau\phi\qty[\partial_\tau^2+\partial_\rho^2]\phi\nonumber\\
&= -\frac{1}{2}\int d\rho d\tau\frac{1}{\sqrt{2\pi}}\int e^{i\rho\lambda_1}\Tilde{\phi}(\tau,\lambda_1)d\lambda_1\qty[\partial_\tau^2+\partial_\rho^2]\frac{1}{\sqrt{2\pi}}\int e^{-i\rho\lambda_2}\Tilde{\phi}^*(\tau,\lambda_2)d\lambda_2\nonumber\\
&=-\frac{1}{2}\int d\tau d\lambda_1 d\lambda_2\qty[\frac{1}{2\pi}\int d\rho e^{i\rho(\lambda_1-\lambda_2)}]\Tilde{\phi}(\tau,\lambda_1)\qty[\partial_\tau^2-\lambda_2^2]\Tilde{\phi}^*(\tau,\lambda_2)\nonumber\\
&=-\frac{1}{2}\int d\tau d\lambda_1 d\lambda_2\delta(\lambda_1-\lambda_2)\Tilde{\phi}(\tau,\lambda_1)\qty[\partial_\tau^2-\lambda_2^2]\Tilde{\phi}^*(\tau,\lambda_2)\nonumber\\
&=-\frac{1}{2}\int d\tau d\lambda\Tilde{\phi}(\tau,\lambda)\qty[\partial_\tau^2-\lambda^2]\Tilde{\phi}^*(\tau,\lambda).\label{acwwlf}
\end{align}

Furthermore, from the Fourier transform \eqref{tfptr}, we have that \(\Tilde{\phi}^*(\tau,\lambda) = \Tilde{\phi}(\tau,-\lambda)\), which makes it evident that the complex field \(\Tilde{\phi}(\tau,\lambda)\) is composed of a real part that is even in \(\lambda\) and an imaginary part that is odd in \(\lambda\). Specifically,
\begin{align}
    \Tilde{\phi}^*(\tau,\lambda) &= \Tilde{\phi}(\tau,-\lambda)\nonumber\\
    \Tilde{\phi}_r(\tau,\lambda) - i\Tilde{\phi}_i(\tau,\lambda) &= \Tilde{\phi}_r(\tau,-\lambda) + i\Tilde{\phi}_i(\tau,-\lambda),\label{cpicc}
\end{align}
where the indices \(r\) and \(i\) denote the real and imaginary parts, respectively.\\

The term $\Tilde{\phi}(\tau, \lambda)\partial_\tau^2 \Tilde{\phi}^*(\tau, \lambda)$ in the action \eqref{acwwlf} can be rewritten as $-\partial_\tau \Tilde{\phi}(\tau, \lambda)\partial_\tau \Tilde{\phi}^*(\tau, \lambda)$ through integration by parts. Thus, in terms of the real and imaginary components, we have
\begin{equation}
    \partial_\tau \Tilde{\phi}(\tau, \lambda)\partial_\tau \Tilde{\phi}^*(\tau, \lambda) = (\partial_\tau \Tilde{\phi}_r(\tau, \lambda))^2 + (\partial_\tau \Tilde{\phi}_i(\tau, \lambda))^2.\label{18.,,}
\end{equation}

Similarly, for the product \(\Tilde{\phi}(\tau, \lambda)\Tilde{\phi}^*(\tau, \lambda)\), in terms of the real and imaginary components, we have
\begin{equation}
    \Tilde{\phi}(\tau, \lambda)\Tilde{\phi}^*(\tau, \lambda) = \Tilde{\phi}_r^2(\tau, \lambda) + \Tilde{\phi}_i^2(\tau, \lambda).\label{18.,,.}
\end{equation}

As we can observe, the terms in \eqref{18.,,} and \eqref{18.,,.} are real quantities, which indicates that it is possible to express them in terms of a real scalar field \(\psi(\tau, \lambda)\) as \(\psi(\tau, \lambda)^2\) and \((\partial_\tau \psi(\tau, \lambda))^2\), respectively:
\begin{align}
    (\partial_\tau \Tilde{\phi}_r(\tau, \lambda))^2 + (\partial_\tau \Tilde{\phi}_i(\tau, \lambda))^2 &= (\partial_\tau \psi(\tau, \lambda))^2,\label{10000-}\\
\Tilde{\phi}_r^2(\tau, \lambda) + \Tilde{\phi}_i^2(\tau, \lambda) &= \psi(\tau, \lambda)^2.\label{10000--}
\end{align}

From the system given by \eqref{10000-} and \eqref{10000--}, we have
\begin{align}
    \psi&=\pm\sqrt{\Tilde{\phi}_r^2 + \Tilde{\phi}_i^2} \to (\partial_\tau \psi)^2=\frac{(\Tilde{\phi}_r\partial_\tau\Tilde{\phi}_r + \Tilde{\phi}_i\partial_\tau\Tilde{\phi}_i)^2}{\Tilde{\phi}_r^2 + \Tilde{\phi}_i^2}.
\end{align}
By equating this expression with the one given for $(\partial_\tau \psi)^2$ in \eqref{10000-}, we obtain
\begin{equation}
    \frac{(\Tilde{\phi}_r\partial_\tau\Tilde{\phi}_r + \Tilde{\phi}_i\partial_\tau\Tilde{\phi}_i)^2}{\Tilde{\phi}_r^2 + \Tilde{\phi}_i^2}=(\partial_\tau \Tilde{\phi}_r)^2 + (\partial_\tau \Tilde{\phi}_i)^2.
\end{equation}

From this, we arrive to
\begin{equation}
    \qty(\frac{\partial_\tau \Tilde{\phi}_i}{\Tilde{\phi}_i}-\frac{\partial_\tau \Tilde{\phi}_r}{\Tilde{\phi}_r})^2=0 \to 
    \partial_\tau( \ln\Tilde{\phi}_i(\tau, \lambda)-\ln\Tilde{\phi}_r(\tau, \lambda))=0,
\end{equation}

Thus, it becomes evident that the difference of the logarithms must be a real function that is either constant or dependent solely on \(\lambda\). However, due to the parity condition \eqref{cpicc}, we have
\begin{equation}
    \ln\Tilde{\phi}_i(\tau, \lambda)=\ln\Tilde{\phi}_r(\tau, \lambda)+\ln f(\lambda),
\end{equation}
and therefore
\begin{equation}
    \Tilde{\phi}_i(\tau, \lambda) = f(\lambda) \Tilde{\phi}_r(\tau, \lambda),
\end{equation}
such that, from \eqref{cpicc}, the real function \(f(\lambda)\) is odd in $\lambda$.\\

Then, from \eqref{18.,,}, \eqref{18.,,.}, \eqref{10000-}, and \eqref{10000--}, we obtain
\begin{align}
    \Tilde{\phi}(\tau, \lambda)\Tilde{\phi}^*(\tau, \lambda) = \Tilde{\phi}_r(\tau, \lambda)^2 (1 + f(\lambda)^2)&=\psi(\tau, \lambda)^2,\\
\partial_\tau \Tilde{\phi}(\tau, \lambda)\partial_\tau \Tilde{\phi}^*(\tau, \lambda) = (\partial_\tau \Tilde{\phi}_r(\tau, \lambda))^2 (1 + f(\lambda)^2)&=(\partial_\tau \psi(\tau, \lambda))^2.
\end{align}
Since \(1 + f(\lambda)^2\) is always positive, we define
\begin{equation}
    \psi(\tau, \lambda) := \Tilde{\phi}_r(\tau, \lambda) \sqrt{1 + f(\lambda)^2}.
\end{equation}
Therefore, the action \eqref{cpicc} takes the following form:
\begin{equation}
    S_E[\psi] = -\frac{1}{2} \int d\tau d\lambda \, \psi(\tau, \lambda) \left[\partial_\tau^2 - \lambda^2 \right] \psi(\tau, \lambda).
\end{equation}

By implementing the discretization \(\lambda \to \lambda_\alpha\), we obtain the harmonic oscillator action for each value of \(\lambda_\alpha\):
\begin{equation}
    S_E[\psi_{\lambda_\alpha}] = -\frac{1}{2} \int d\tau \, \psi_{\lambda_\alpha}(\tau) \left[\partial_\tau^2 - \lambda_\alpha^2 \right] \psi_{\lambda_\alpha}(\tau),
\end{equation}
where $\psi_{\lambda_\alpha}(\tau)=\psi(\tau, \lambda_\alpha)\sqrt{\Delta\lambda} = \Tilde{\phi}_r(\tau, \lambda_\alpha) \sqrt{1 + f(\lambda_\alpha)^2}\sqrt{\Delta\lambda}$, and notice that the interval \(\Delta\lambda = \lambda_{\alpha+1} - \lambda_\alpha>0\).\\

The path integral for the action \eqref{acwwlf} in its discrete form in $\lambda$ is proportional to
\begin{equation}
    \prod_{\alpha}\int D\psi_{\alpha}e^{S_E[\psi_{\lambda_\alpha}]}.
\end{equation}

As mentioned in section \ref{secyl}, a fundamental requirement for the applicability of the Laflamme method is the periodicity in \(\tau\), given by \(\tau \sim \tau + \beta\). Therefore, we can express the path integral as the product of two parts: one corresponding to \(0 < \tau < \beta/2\) and the other to \(\beta/2 < \tau < \beta\). This becomes evident if we also discretize in \(\tau\), which is well known (see, for example, \cite{Peskin:1995ev}), so that we ultimately arrive at
\begin{equation}
P_{\alpha} \propto \prod_\alpha \int_{\mathscr{C}_+} D\psi_{\lambda_\alpha} e^{-S_E[\psi_{\lambda_\alpha}]} \prod_\alpha \int_{\mathscr{C}_-} D\psi_{\lambda_\alpha} e^{-S_E[\psi_{\lambda_\alpha}]}, \label{pdis}
\end{equation}
where \(\mathscr{C}_\pm\) denotes the class of regular fields defined on the background geometry \(M_\pm\), with \(M_+\) corresponding to \(0 < \tau < \beta/2\) and \(M_-\) corresponding to \(\beta/2 < \tau < \beta\).

\section{Relation between \(\tau\) and \(t_E\) for the causal diamond}\label{t-m.,k}
To calculate the value of \(\beta\) associated with the causal diamond, we need to express the coordinates in this system, namely $(\tau, \rho)$, in terms of the coordinates in the Euclidean plane \((t_E, x)\). From Eqs.~(\ref{cdtrrm}) and (\ref{tdxdtrxrtauxi}, we can write $\eta$ and $\xi$ in terms of $t_d$ and $x_d$ as :
\begin{align}
    \eta &= \alpha \tanh^{-1}\left( \frac{4\alpha^2 t_{\!_d}}{4\alpha^2(x_{\!_d}+\alpha) - 2\alpha[(x_{\!_d}+\alpha)^2-t_{\!_d}^2]}\right) \label{eq:diamond-jacobson} \\
    \xi &= \alpha \ln\left( \frac{16\alpha^4 t_{\!_d}^2 - \{ 4\alpha^2(x_{\!_d}+\alpha) - 2\alpha[(x_{\!_d}+\alpha)^2-t_{\!_d}^2] \}^2}{\alpha^2[(x_{\!_d}+\alpha)^2 - t_{\!_d}^2 - 4\alpha x_{\!_d}]^2} \right) 
    \; .
\end{align}

From this we can get the Euclideanized version by using the definition $t_d = -it_E$ and $\eta = -i\tau$
\begin{equation}
    \tau = \alpha \tan^{-1}\left( \frac{4\alpha^2 t_{E}}{4\alpha^2(x_{\!_d}+\alpha) - 2\alpha[(x_{\!_d}+\alpha)^2 + t_{E}^2]}\right)
\end{equation}

Thus, for a static observer situated at $x = 0$, $\tau$ reduces to
\begin{equation}
    \tau(t_E, 0) = \alpha \tan^{-1}\left(\frac{2 \alpha t_E}{\alpha^2 - t_E^2}\right)\;. \label{valjsl09/}
\end{equation}

\section{Thermofield double state in Rindler geometry}\label{anlysnrin}
Knowing that the metric of the causal diamond is conformal to Rindler, as shown in \eqref{len56}, the Euclidean action for Rindler in coordinates \(\qty(\tau,\rho)\) will indeed be the same as that given in \eqref{swphi}. Thus, imposing periodicity in the coordinate \(\tau\), i.e., \(\tau \sim \tau + \beta\), it now corresponds to the periodic coordinate of the Laflamme cylinder. Therefore, the equation \eqref{psi+++} in this context will be
\begin{equation}
    \Psi_+[\psi_1,\psi_2] = \bra{\psi_2}e^{-\frac{\beta}{2}H}\ket{\psi_1}.\label{amlty31}
\end{equation}

From the coordinate transformation \eqref{cdtrrm}, with $\xi=\alpha e^{\rho/\alpha}$, we have that the generator of time evolution in Rindler spacetime $\qty(\eta,\rho)$ corresponds to the boost generator in the \(x\) direction of Minkowski spacetime $\qty(t,x)$ up to a factor of \(1/\alpha\):
\begin{equation}
    \partial_{\eta} = \frac{\partial x}{\partial\eta}\partial_x + \frac{\partial t}{\partial\eta}\partial_t = \frac{1}{\alpha}(t\partial_x + x\partial_{t}).\label{kieta}
\end{equation}

\begin{figure}[t]
    \centering
    \begin{subfigure}{0.30\textwidth}
        \includegraphics[width=\linewidth]{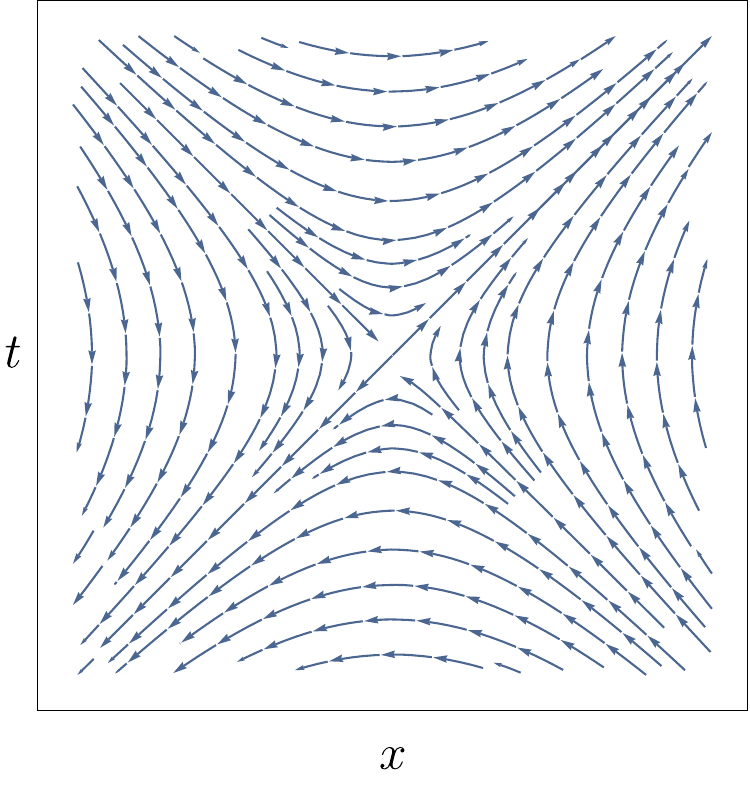}
\captionsetup{justification=centering} 
        \caption{}
        \label{fig5a}
    \end{subfigure}
    \hfill
    \begin{subfigure}{0.3\textwidth}
        \includegraphics[width=\linewidth]{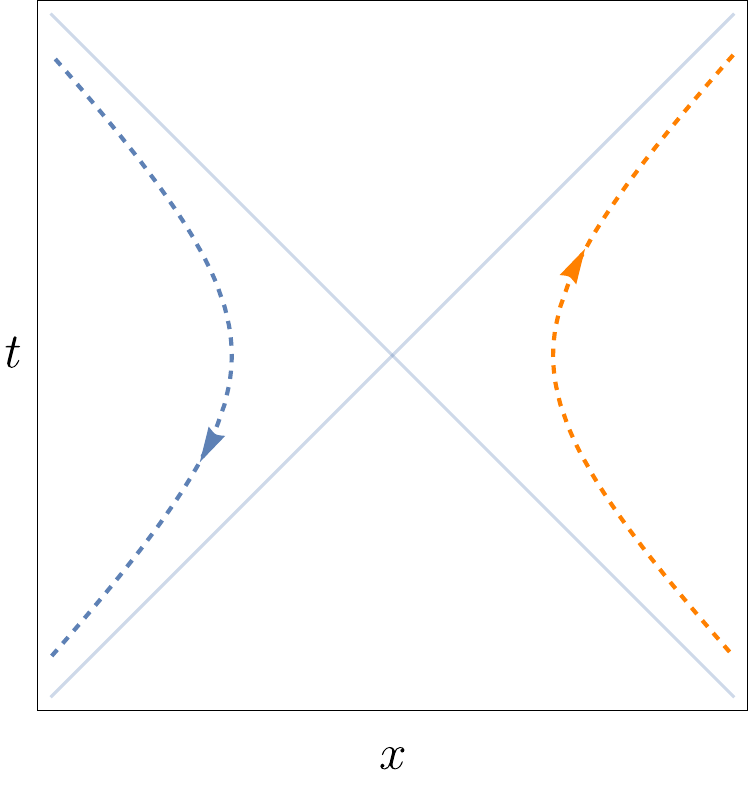}
\captionsetup{justification=centering}
        \caption{}
        \label{fig5b}
    \end{subfigure}
    \hfill
    \begin{subfigure}{0.3\textwidth}
        \includegraphics[width=\linewidth]{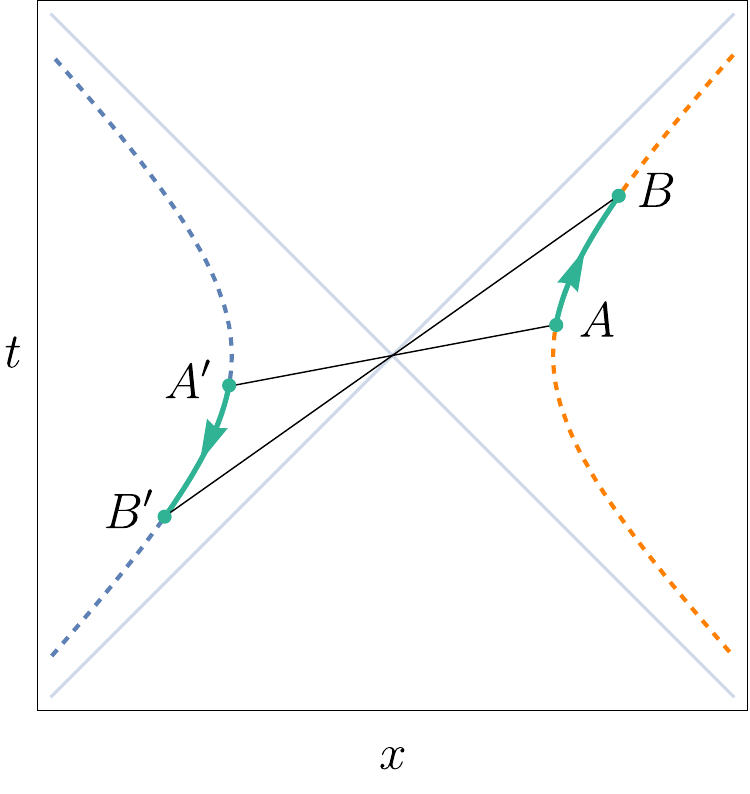}
\captionsetup{justification=centering}
        \caption{}
        \label{fig5c}
    \end{subfigure}    
    \caption{(a) Integral curves of the vector field $\alpha\partial_\eta$ on Minkowski spacetime.\\ (b) Trajectories for a fixed value of $\rho$. (c) Evolution from $A$ to $B$ related by parity and time-reversal to that from $A'$ to $B'$.}
    \label{fig}
\end{figure}

The integral curves of this vector field are shown in figure \ref{fig5a}, from this, we observe that the temporal evolution in the right wedge is future-directed, while in the left wedge, it is past-directed when taking the temporal evolution in Minkowski spacetime as the reference. Similarly, it is important to mention that the trajectories described by the observer in Rindler spacetime across the entire range of $\eta$ values, are given by the semi-hyperbolas in the right and left wedges, such that together they complement the hyperbola defined over the entire Minkowski spacetime (see figure \ref{fig5b}), whose asymptotes are given by the null horizons $t=\pm x$. Furthermore, the temporal evolution on one of the semi-hyperbolas automatically describes the other through parity and time-reversal transformations in $\qty(t,x)$, as we note from figure \ref{fig5c}.\\

\begin{figure}[t]
    \centering
    \begin{subfigure}{0.29\textwidth}
        \includegraphics[width=\linewidth]{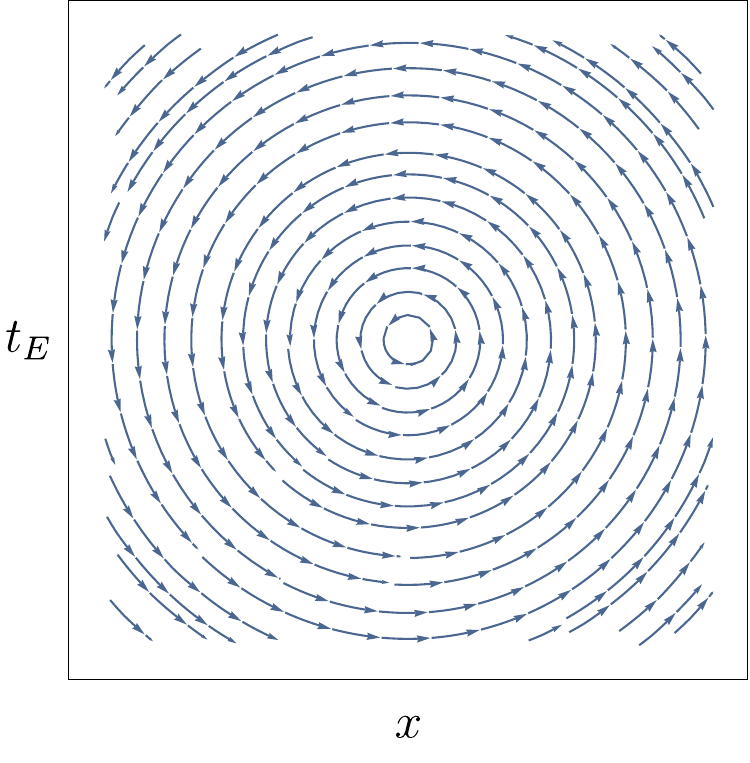}
\captionsetup{justification=centering} 
        \caption{}
        \label{fig6a}
    \end{subfigure}
    \hspace{.5cm}
    \begin{subfigure}{0.29\textwidth}
        \includegraphics[width=\linewidth]{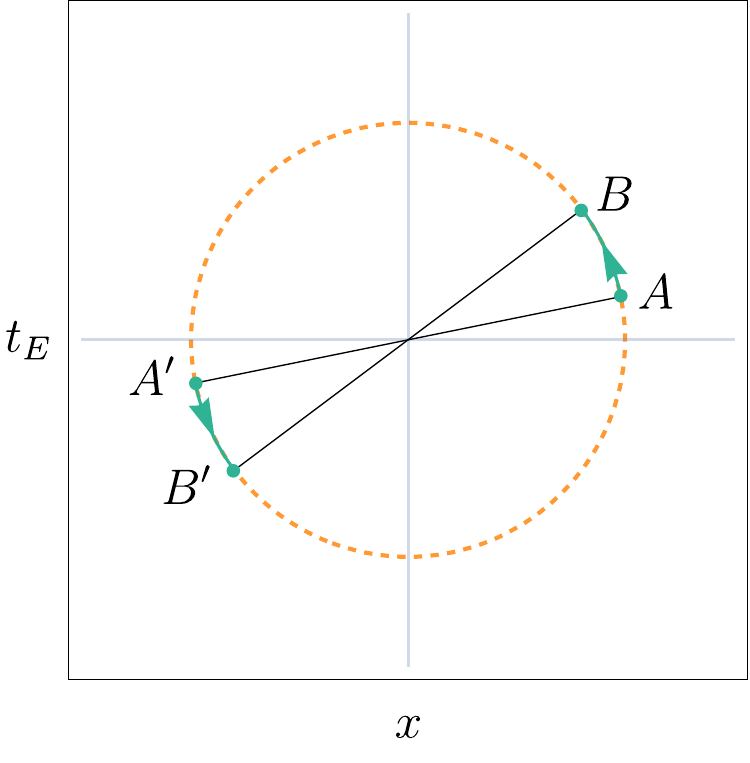}
\captionsetup{justification=centering}
        \caption{}
        \label{fig6b}
    \end{subfigure} 
\hspace{.5cm}
\begin{subfigure}{0.29\textwidth}
        \includegraphics[width=\linewidth]{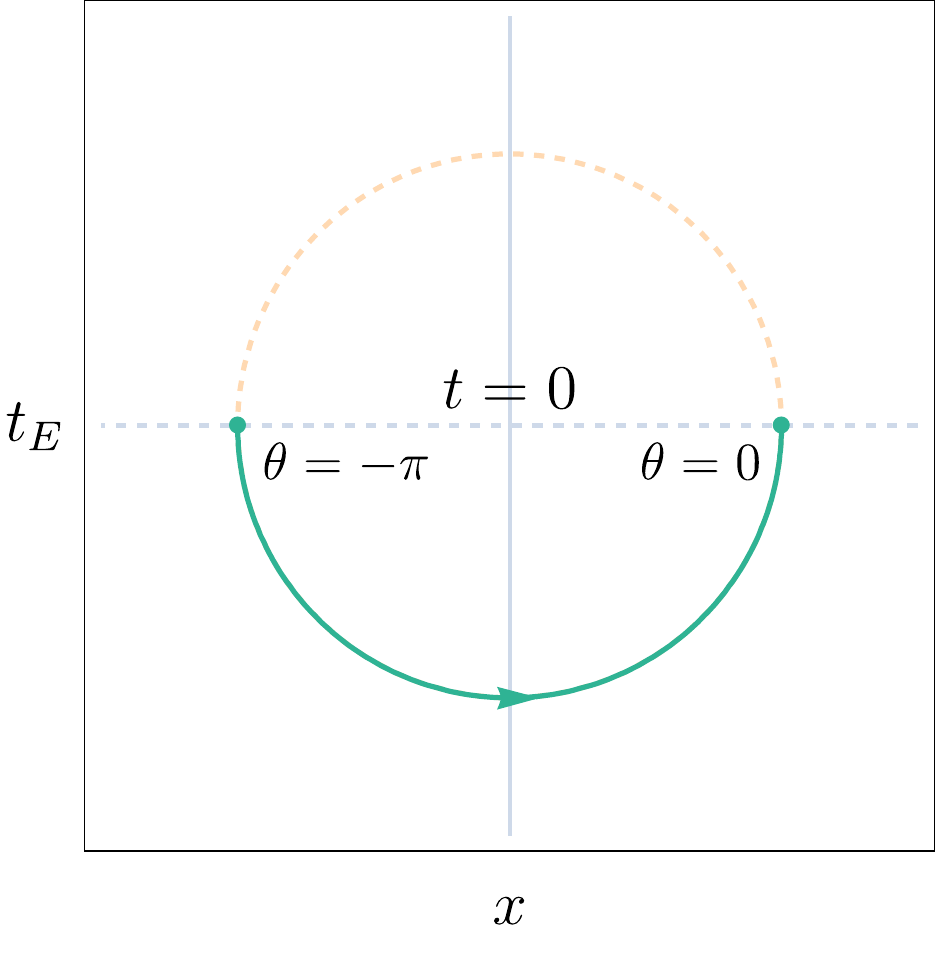}
\captionsetup{justification=centering}
        \caption{}
        \label{fig6c}
    \end{subfigure}
    \caption{(a) Integral curves of the vector field $\alpha\partial \tau$ in the Euclidean plane.\\ (b) Evolution from $A$ to $B$ related by parity and (Euclidean) time-reversal to that from $A'$ to $B'$. (c) Evolution by $\pi$ radians in the lower half of the Euclidean plane.}
    \label{fig}
\end{figure}

Returning to the Euclidean signature ($\eta=-i\tau$ and $t = -it_E$), the vector field \eqref{kieta} becomes the generator of counterclockwise rotations in the Euclidean plane up to a factor of $1/\alpha$  (see figure \ref{fig6a}),
\begin{equation}
\partial_{\tau}=\frac{1}{\alpha}(-t_E\partial_x + x\partial_{t_E}),\label{orpetex}
\end{equation}
As a consequence of the Wick rotation, we have different trajectories compared to those seen in the Lorentzian case. The geometric flow generated by \(\partial_\tau\) is related to rotations in the Euclidean plane by \(\tau = \alpha \theta\), where \(\theta\) is an angular coordinate with periodicity \(2\pi\). Furthermore, the trajectories obtained across the entire range of $\tau/\alpha$ values do not complement each other in the sense we saw in the Lorentzian case, where both semi-hyperbolas formed a whole in Minkowski spacetime. Here, for both coordinate transformations in the left and right wedges, we obtain a circumference in the entire Euclidean plane, which indicates that this mapping is two-to-one. However, complementarity is observed when analyzing $\tau/\alpha$ evolution, as one trajectory immediately defines the other, again through parity and time-reversal transformations in $\qty(t_E,x)$ (which in this case would also be parity in $t_E$), as shown in figure \ref{fig6b}.\\

From \eqref{orpetex}, we have that the vector field \(\alpha\partial_\tau\) is the generator of rotations in the Euclidean plane described by the coordinates \(\qty(t_E, x)\). We know that rotations in this plane have a periodicity of \(2\pi\), so we can associate the vector field \(\alpha\partial_\tau\) with the angular displacement generator denoted by \(\partial_\theta\). Thus, the transition amplitude \eqref{amlty31}, in terms of the Hamiltonian operator associated with \(\partial_\theta\), given by \(H_\theta = \alpha H\), will be
\begin{equation}
    \Psi_+[\psi_1,\psi_2] = \bra{\psi_2}e^{-\frac{\beta}{2}H}\ket{\psi_1}= \bra{\psi_2}e^{-\qty(\frac{\beta}{2\alpha})H_\theta}\ket{\psi_1}=\bra{\psi_2}e^{-\pi H_\theta}\ket{\psi_1},\label{asdk34}
\end{equation}
from which we see that \(\beta/(2\alpha)\) is precisely half of the period in \(\theta\), which corresponds to \(\pi\). Therefore, we find that \(\beta\) is given by
\begin{equation}
    \beta=2\pi\alpha.
\end{equation}
This result corresponds to the temperature observed in the Unruh effect \cite{Unruh:1976db,Davies:1976ei,Davies:1976hi,Davies:1977bgr, Davies:1974th}, \(T=a/(2\pi)\), where \(a=\alpha^{-1}\) is the constant acceleration of the non-inertial observer in Minkowski spacetime.\\

From the right-hand side of \eqref{asdk34}, we observe that the propagation of the state \(\ket{\psi_1}\) to \(\ket{\psi_2}\) in the Euclidean plane \(\qty(t_E,x)\) corresponds to an angular evolution of \(\pi\). Specifically, we consider this propagation to occur from \(\theta=-\pi\) to \(\theta=0\), as shown in figure \ref{fig6c}. The purpose of this choice is due to the fact that Lorentzian and Euclidean time coincide at \(t=t_E=0\), and since the \(x\) coordinate is unaffected by the Wick rotation, the states defined on the line \(t=t_E=0\) can be interpreted as those on Minkowski spacetime. Additionally, we know that the states \(\ket{\psi_{1,2}}\) can be expressed in the eigenbasis of the Hamiltonian operator in Rindler: \(H=\alpha^{-1} H_\theta\). For this, we must define which wedge of Rindler spacetime the real scalar field is in, from whose Euclidean action the field configurations on its boundary, i.e., \(\ket{\psi_{1,2}}\), are derived. Let us consider, then, that the field was defined in the right wedge (if we choose the left wedge, the procedure is the same), so the operator \(H\) has the energy spectrum for the field states in the right wedge of Rindler. Thus, both \(\ket{\psi_1}\) and \(\ket{\psi_2}\) have this nature:
\begin{equation}
    \ket{\psi_1}=\ket*{\Tilde{\psi}_R},\quad \ket{\psi_2}=\ket{\psi_R}\quad\Rightarrow \quad\psi_1=\Tilde{\psi}_R,\quad \psi_2=\psi_R,\label{0if36}
\end{equation}
where the tilde is only used to distinguish between the states. However, given the angular propagation of \(\pi\) for these states, it means that one of them is defined on the left side of the Euclidean plane, while the other is on the right, both situated at \(t_E=0\). In other words, the initial state in this propagation belongs to the left wedge of Rindler, but that would not be possible since the Hamiltonian operator could not act on that state. For this reason, we recognize in \eqref{0if36} that the state \(\ket*{\Tilde{\psi}_R}\) must be given by
\begin{equation}
    \ket*{\Tilde{\psi}_R}=\Theta\ket{\psi_L},
\end{equation}
where \(\Theta\) is the anti-unitary \(CPT\) operator, whose necessity arises because the Hamiltonian operator will act only on the states defined in the right wedge of Rindler. Recall that both wedges are related through parity and time-reversal operations. Note that the action of the anti-unitary operator allows for a natural map between the Hilbert spaces for the fields on the left $(x<0)$ \(\mathcal{H}_L\) and on right $(x>0)$ \(\mathcal{H}_R\) \cite{Harlow:2014yka}:
\begin{equation}
\Tilde{\psi}_R=\Theta\psi_L\Theta^{-1},\quad \ket*{\Tilde{\psi}_R}=\Theta\ket{\psi_L}.\label{mimeq616}
\end{equation}
From \eqref{0if36} and \eqref{mimeq616} we identify $\psi_1=\Theta\psi_L\Theta^{-1}$. In addition, it is worth mentioning that, since this is a real scalar field, the charge conjugation does not have any effect.\\

Given all of the above, we have that the transition amplitude \eqref{asdk34} is given by
\begin{equation}
\Psi_+\qty[\psi_L,\psi_R]=\bra{\psi_R}e^{-\pi H_\theta}\Theta\ket{\psi_L},\label{frtfl}
\end{equation}

Let us insert the operator \(1=\sum_n\ket{n_R}\bra{n_R}\), where \(\ket{n_R}\) are the eigenbasis of the Hamiltonian in the right wedge:
\begin{align}
    \Psi_+\qty[\psi_L,\psi_R]&=\sum_n\bra{\psi_R}\ket{n_R}\bra{n_R}e^{-\pi\alpha H}\Theta\ket{\psi_L}\nonumber\\
    &=\sum_n e^{-\pi\alpha E_n}\bra{\psi_R}\ket{n_R}\bra{\psi_L}\Theta^{-1}\ket{n_R},
\end{align}
where we have used \(\bra{n_R}\Theta\ket{\psi_L} = \bra{\psi_L}\Theta^{-1}\ket{n_R}\) (see Appendix E in \cite{Valdivia-Mera:2020nko}).\\

Thus, the transition amplitude \eqref{frtfl} takes the form of the following inner product:
\begin{align}
\Psi_+\qty[\psi_L,\psi_R]&=\bra{\psi_R}e^{-\pi H_\theta}\Theta\ket{\psi_L}\nonumber\\
&=\bra{\psi_L,\psi_R}\qty(\sum_n e^{-\pi\alpha E_n}\ket{n_R}\otimes\Theta^{-1}\ket{n_R})\propto \bra{\psi_L,\psi_R}\ket{0(\beta)},\label{sdaesd41}
\end{align}
such that the normalized ket on the right-hand side of \eqref{sdaesd41} is given by
\begin{equation}
    \ket{0(\beta)} = \frac{1}{\sqrt{Z(\beta)}} \sum_{n=0}^{\infty} e^{-\frac{\beta}{2}E_n} \ket{n_R} \otimes \Theta^{-1} \ket{n_R},\label{tfdsrej}
\end{equation}
where we have used the fact that \(\beta/2 = \pi \alpha\).\\

The state $\ket{0(\beta)}$ corresponds to the Minkowski vacuum introduced by Umezawa and Takahashi, commonly referred to as the thermofield double state \cite{Takahasi:1974zn, Umezawa:1982nv, Matsumoto:1982ry, Takahashi:1996zn}. It is described as a direct product of the Hamiltonian eigenbasis of the field in the right and left Rindler wedges.

\newpage
\section*{References}
\bibliography{biblio}

\end{document}